\documentclass[aps,prx,reprint,superscriptaddress,footnote,nofootinbib, twocolumn,notitlepage]{revtex4-2}
\usepackage{amsmath, amsthm, amssymb,amsfonts,mathbbol,amstext}
\usepackage{braket}
\usepackage{pifont}
\usepackage{graphicx}
\usepackage{dcolumn}
\usepackage{bm}
\usepackage{bbm}
\usepackage{hyperref}
\usepackage{mathtools}
\usepackage{comment}
\usepackage{color}
\usepackage{multirow}
\usepackage{diagbox}
\usepackage{float}
\usepackage{xcolor}
\usepackage{makecell}
\usepackage[normalem]{ulem}
\usepackage{soul}
\usepackage{tikz}
\usepackage{booktabs}
\usepackage[inline]{enumitem}
  \newcommand{\Kc}[0]{{\ensuremath K_c}}
\newcommand{\su}[0]{{\mathfrak{su}}}

\newtheorem{proposition}{Proposition}
\theoremstyle{definition}
\newtheorem{definition}{Definition}

\usepackage[english]{babel}

\usepackage[letterpaper,top=2cm,bottom=2cm,left=3cm,right=3cm,marginparwidth=1.75cm]{geometry}

\begin{document}
\title{%
Look Mum, No Hands! Quantum State Preparation without Time-Dependent Control%
}

\author{T. Andrzejewski}
\affiliation{Atominstitut, Technische Universit{\"a}t Wien, Stadionallee 2, 1020 Vienna, Austria}

\author{Y. Minoguchi}
\affiliation{Atominstitut, Technische Universit{\"a}t Wien, Stadionallee 2, 1020 Vienna, Austria}

\author{P. Rembold}
\affiliation{Atominstitut, Technische Universit{\"a}t Wien, Stadionallee 2, 1020 Vienna, Austria}

\author{K. Szyma\'nski}
\email{k.sz@quantumstat.es}
\affiliation{Research Center for Quantum Information, Slovenská Akadémia Vied, Dúbravská cesta 9, 84511 Bratislava, Slovakia}
\affiliation{Naturwissenschaftlich-Technische Fakultät, Universität Siegen, Walter-Flex-Straße 3, 57068 Siegen, Germany}

\begin{abstract}
How much can be prepared with a quantum system whose Hamiltonian cannot be modulated in time? Common intuition suggests that the answer is not much. Without a knob to turn during the evolution, a system seems confined to a fixed trajectory through its state space, while time-dependent control can steer it almost anywhere. We put this intuition to the test by studying autonomous control, i.e., state preparation with a time-invariant Hamiltonian, where the only freedom lies in the static coupling strengths and the evolution time. We characterize basic mathematical properties of the autonomously reachable set and introduce two criteria of unreachability: a certificate that is cheap to evaluate and rigorously excludes that a target state can be reached for any choice of control parameters, and a stronger criterion based on exhaustive numerical optimization. With these tools we estimate the probability that a typical state in Hilbert space is unreachable under randomly drawn control terms. The dimensionality of the reachable set turns out to be independent of the system, yet the fraction of Hilbert space lying close to it depends strongly on the choice of control Hamiltonians. Strikingly, for sufficiently random controls and long enough evolution times, almost every state can become approximately reachable, even with the fewest generators that permit full controllability with time-dependent couplings. 
{Autonomous control can approach the expressivity of time-dependent control, at the cost of longer evolution times.}
\end{abstract}

\maketitle

\section{Introduction}
{Given a set of controls, what fraction of quantum states can we access without time dependence?
The broader question of controllability is central to exploit any of the advantages promised by quantum technology~\cite{Glaser2015,koch_quantum_2022}.}

{Using \textit{time dependent} controls, this question can be answered straightforwardly for closed quantum systems~\cite{DAlessandro2021}.}
Here, control can be considered from the quantum computing perspective, taking the form of sequentially applied gates~\cite{NielsenChuang2010}, or the hardware perspective, where time-dependent Hamiltonians represent electro-magnetic fields that can be applied to the system~\cite{Rembold2020}. 
The additional level of complexity introduced by time-dependence comes with opportunities and complications alike.
Fast pulses may, for example, lead to unwanted frequency components~\cite{motzoi_simple_2009}, they require slicing up the time-ordered evolution for modelling~\cite{blanes_magnus_2009}, but also allow for shortcuts to adiabaticity and a plethora of other control methods that improve the robustness or speed of quantum protocols~\cite{koch_quantum_2022}.

{Here we ask how much control is left without time-dependence.}
More precisely, how many time-independent controls are required to explore the full state space from a given initial state?
{When this is not met, can we identify the states that stay out of reach?}
Our results provide insights into when time-independent control may be enough.

\begin{figure*}[t]
\includegraphics[width=.7\textwidth]{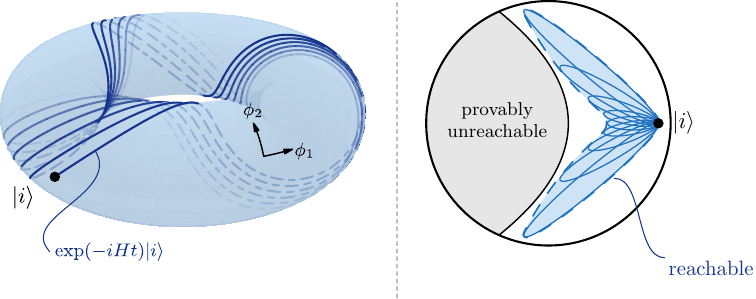}
\caption{Schematic depiction of autonomous reachability. Left:~A fixed Hamiltonian leads to an orbit $\exp(-iH t)\lvert i\rangle$, which for generic $H$ is dense on a torus of the eigenstates' phases, so that the one-dimensional orbit fills a $d$-dimensional torus.
Right: As the Hamiltonian parameters are varied, the torus changes, sweeping a higher-dimensional set. Under the right conditions, for $K$ parameters, a set of $K+d-1$ dimensions is covered (counting the global phase; Sec.~\ref{sec:theory}). Relaxation of the strict reachability criteria leads to the variance criterion (Sec. \ref{sec:variance}), ascertaining when a particular state cannot be reached with a particular Hamiltonian family from a fixed state. }
\label{fig:concept}
\end{figure*}

{Mathematically, how does time independence affect control complexity?}
Time-dependent control exploits the commutators generated when switching between Hamiltonians. 
Hence, state reachability is straightforwardly characterised by the dynamical Lie algebra of the generators~\cite{JurdjevicSussmann1972, albertini_notions_2003}. 
The same mathematical methods do not apply to time-independent Hamiltonians, creating the need for a different reachability analysis.
Still, the dynamics of freely evolving systems are far from trivial: 
for random geometrically two-local Hamiltonians, for example, approximating the output probabilities of the resulting dynamics is likely to be classically intractable~\cite{quek2025quantumadvantagerandomgeometricallytwolocal}. 
{Indeed, quantum computation can in general be formulated time-independently following the original Feynman-Kitaev Hamiltonians~\cite{Feynman1986,Kempe2006}.}
Moreover, any piecewise-constant time-dependent control protocol can be translated into a time-independent version by embedding the system in a larger one~\cite{mooney_time_2025II}, suggesting a similar expressivity at the cost of system size.

{Let us explore the existing application of time-independent control---called autonomous from here on}. 
Its relative experimental simplicity is, e.g., exploited by analogue quantum simulators studying quench dynamics~\cite{bernien2017,daley2022}.
It does not require specialised components for high-frequency modulation and reduces control noise.
Indeed, autonomous controls have been investigated to generate entanglement~\cite{cieslinski_fastest_2023} and for state transfer~\cite{Bose2003,Christandl2004,Kay2010}. In the latter case, autonomous controls can lead to remarkable robustness against Hamiltonian uncertainties~\cite{schirmer_design_2018,weidner_energy_2025}.

{In a computational context, 
the added simplicity of autonomous control could provide an advantage, e.g., for variational quantum algorithms.}
They translate complex problems into Hamiltonians, whose ground states encode the solution.
Unfortunately, the classical optimisers employed to explore the state space tend to stall when the complex controls result in so-called barren plateaus~\cite{holmes_connecting_2022,Larocca2023_1}, where vanishing gradients hinder further improvements.
Recently, Tiwari et al. suggested using autonomous controls~\cite{tiwari_time_efficient_2025}, optimising the weights of the generators of a single constant Hamiltonian, rather than a full time-dependent protocol to avoid barren plateaus.
Whether autonomous Hamiltonians provide a viable alternative on bigger scales depends on their expressivity.
This raises the larger question, how to determine whether an unknown target state is reachable.

From a fundamental perspective, we can translate the question of time-independent reachability into the context of a toy model for a time-independent universe. 
After all, the laws of physics themselves are empirically time-independent as evidenced by energy conservation~\cite{Noether1918}. In our pocket-sized universe, each Hamiltonian term is a law and its weight a constant of nature. 
How rich must these laws be for every configuration to lie within reach for some given constants of nature, with nobody continuously turning the knobs?

A target state is called \textit{reachable}, if it can evolve from an initial state  given some choice of couplings and waiting time.
We want to understand the general behaviour of autonomous control, thus considering many randomised control terms and targets, as opposed to investigating a single fine-tuned example.
{The questions raised so far can thus be formalized}: \begin{enumerate*}[label=(\alph*)]
\item for a given set of control terms, which targets remain out of reach however the couplings and the waiting time are chosen? And  \item how many control terms are needed before a generic target comes within reach? \end{enumerate*}

The two questions lead to two independent results. For the first (see Sec.~\ref{sec:variance}), we find a simple certificate of unreachability that requires no optimization at all. Autonomous evolution conserves the energy statistics of the state, and these statistics are determined by the means and covariances of the control terms. If no choice of couplings can reconcile these statistics between the initial and the target state, the two are not connected by any autonomous evolution. 

For the second question (see Sec.~\ref{sec:spectral_crit}), we compare the populations of the initial and the target state in the energy eigenbasis of the Hamiltonian beyond the first two moments. Autonomous evolution leaves these populations untouched and only advances the relative phases between energy eigenstates. Matching populations is therefore necessary for reachability, and for rationally independent energy gaps it is also sufficient: the phases then wind around without ever repeating and eventually come arbitrarily close to any desired configuration illustrated in Fig.~\ref{fig:concept}. Reachability thus reduces to whether the couplings can be tuned such that initial and target state share the same populations. Since this requires tuning populations alone rather than populations and phases, autonomous control may be at a smaller disadvantage than its reduced expressivity suggests. 

Whether this holds depends on the details of the control, i.e., the algebra of the control Hamiltonians (see Sec.~\ref{sec:comparison_DLA}). For multi-qubit Pauli strings, time-dependent control becomes universal with a number of control terms that grows only linearly with the number of qubits.
However, the number of required control terms for autonomous control is equal to the Hilbert space dimension. Thus, it grows exponentially with the number of qubits. A different example is given by the generalized Pauli operators of a single qudit. Here, both forms of control require the same number of control terms, equal to the dimension of the qudit. Still, the autonomous evolution may require long waiting times before the phases approach their target.  {Interestingly, we find that the entire state space could be reached with a fidelity of 99\% even for the smallest control sets -- that is, every state lies within some small distance to the strictly autonomously reachable set.} 

Thus, autonomous evolution can match the performance of time-dependent control up to a small accuracy trade-off and longer evolution times.

\section{Setup}
\label{sec:setup}

\subsection{Hamiltonian pools}
\label{subsec:ham_families}

Throughout this work the control terms or generators are drawn from one of two fixed operator \emph{pools}. 
The first is the \emph{qudit pool}: the elementary Hermitian basis on a Hilbert space of dimension $d$, namely the $\frac{d(d-1)}{2}$ off-diagonal symmetric and antisymmetric elements, $X_{jk}$ and $Y_{jk}$, and the $d-1$ traceless diagonal differences $Z_j$, resulting in $d^2-1$ operators in total:
\begin{equation}\begin{aligned}
    X_{jk} & =  \ket{j}\bra{k}+\ket{k}\bra{j}, \\
    Y_{jk} & =  i(\ket{j}\bra{k}-\ket{k}\bra{j}), \\
    Z_j & =  \vert j \rangle \langle j \vert - \vert j+1 \rangle\langle j+1 \vert,\end{aligned}
    \label{eq:quditpool}
\end{equation}
with $0 \le j < k \le d-1$ for $X_{jk}$ and $Y_{jk}$ and $0 \le j \le d-2$ for $Z_j$.

The second, defined for $n$ qubits with dimension $d=2^n$, is the \emph{Pauli pool}: all non-identity Pauli strings, consisting of the Pauli operators $X,Y$ and $Z$, and identity $\mathbb{1}$ on $n$ qubits (which yields $d^2-1=4^n - 1$ operators when excluding the identity), that is 
\begin{equation}
    \{ P_1 \otimes \ldots \otimes P_n : P_i \in \{X,Y,Z,\mathbb{1}\}\} \setminus \{\mathbb{1}_n\}.
\end{equation}
Note that this set of Pauli string generators is another basis which spans the entire Lie algebra $\mathfrak{su}(d)$.

Both pools, in their entirety, span the set of traceless Hermitian operators, and exponentials of their weighted sums can produce any special unitary operator. In other words, access to all of them makes the system fully controllable. 
In the subsequent text we use \emph{control sets}, which are operators $\{h_1, \ldots, h_K\}$ sampled uniformly at random from the chosen pool, subject to the constraint that \emph{time-dependent control could cover the entire state space}. This constraint {is further described in Sec.~\ref{sec:dla} and is} employed to have a direct comparison of our reachability criteria with time-dependent reachability, as otherwise, states found not reachable could be due to the fact that the Hamiltonian family is not expressive enough to even reach the state in the more general case.

For one of the reachability criteria (Sec. \ref{sec:breakdown}), we further restrict the Hamiltonian families so that they have no underlying eigenvalue symmetries; this is done to make sure that states found reachable by the numerical method indeed are close to the reachable set.
\subsection{Reachability}\label{subsec:reachability}

We consider a time-independent (autonomous) version of a question that is common in quantum control: whether a prescribed target state can be reached from a given initial state without any explicit time dependence in the control protocol. 
We start by fixing an initial state $\vert i \rangle$ and a final (target) state $\vert f \rangle$. 

In addition, we are given a set of $K$ Hamiltonian control terms, or Hermitian generators, $\{h_\ell\}_{\ell=1}^K$ with $h_\ell = h_\ell^\dagger$. 
The objective in autonomous control is to find time-independent control parameters or coupling constants $\{\lambda_\ell\}_{\ell=1}^K$ and a suitable duration of the evolution $t_*$ such that
\begin{equation}\label{eq:base_setting}
\vert f \rangle \approx \vert \psi(t_*)\rangle =  e^{-it_* H(\vec{\lambda})} \vert i \rangle,
\end{equation}
within prescribed accuracy or exactly.
For notational convenience, we collect all coefficients into the vector $\vec{\lambda} = (\lambda_1,\ldots,\lambda_K)^\top$.
Note that the evolved state depends on $t_*$ and $\vec{\lambda}$ only through the product $t_*\vec{\lambda}$: rescaling the couplings merely rescales the evolution time, a redundancy we will exploit repeatedly below.
The approximate equality in Eq.~\eqref{eq:base_setting} is given a precise meaning by the following closure-sense definition, which we use throughout.

\begin{definition}[Autonomous reachability]\label{def:reach}
Given an initial state $\vert i \rangle$ and a control set
$\{h_\ell\}_{\ell=1}^K$, a target $\vert f \rangle$ is
\emph{autonomously reachable} if for every distance $\varepsilon > 0$ there exist
couplings $\vec{\lambda}$ and a time $t_*$ such that
\begin{equation}
\label{eq:distbound}
\bigl\Vert\, \vert f \rangle - e^{i\theta} e^{-it_* H(\vec{\lambda})} \vert i \rangle \,\bigr\Vert < \varepsilon,
\end{equation}
for some global phase $\theta$. This is equivalent to $\vert f \rangle$ lying in
the closure of the union of the orbits of $\vert i \rangle$, up to a
global phase.
\end{definition}

The closure is crucial. As we will see in Sec.~\ref{sec:theory}, a
single orbit is a one-dimensional curve, yet the set of states it comes
arbitrarily close to is typically much larger, and it is this larger set
that autonomous control can explore (Fig.~\ref{fig:concept}).

To determine whether $\vert f \rangle$ is reachable from $\vert i \rangle$ given the controls $\{h_\ell\}_{\ell=1}^K$, we will employ the spectral decomposition of the Hamiltonian,
\begin{equation}\begin{aligned}
H(\vec{\lambda})  & =   \sum_{\ell=1}^K \lambda_\ell h_\ell\\
&= \sum_{n=1}^{d} \omega_n(\vec{\lambda}) \vert n_{\vec{\lambda}}\rangle \langle n_{\vec{\lambda}}\vert .
\end{aligned}
\end{equation}

We thus regard the eigenbasis $\{\vert n_{\vec{\lambda}}\rangle\}_{n=1}^d$ and the corresponding eigenvalues $\{\omega_n(\vec{\lambda})\}_{n=1}^d$ as functions of the control parameters $\vec{\lambda}$. 
Let us now express the amplitudes of the initial and target states in the eigenbasis of $H(\vec\lambda)$:
\begin{equation}\label{eq:polar}
\langle n_{\vec{\lambda}} \vert \alpha \rangle
= e^{i\varphi_{\alpha,n}(\vec{\lambda})}\sqrt{p_{\alpha,n}(\vec{\lambda})},
\end{equation}
where $\alpha \in \{i,f\}$. 
Here, $p_{\alpha,n}(\vec{\lambda})$ denotes the population of the eigenstate $\vert n_{\vec{\lambda}}\rangle$ in $\vert \alpha \rangle$, and $\varphi_{\alpha,n}(\vec{\lambda})$ is the corresponding phase. 

With these ingredients we can say that for Definition \ref{def:reach} to hold, there must be a combination of $\vec{\lambda}$, $t_*$, and $\theta$ such that, for all $n$ the system of equations
\begin{eqnarray}
p_{i,n}(\vec{\lambda}) & \approx & p_{f,n}(\vec{\lambda}) \label{eq:distribution} \\
\varphi_{f,n}(\vec{\lambda}) & \approx & 
[\varphi_{i,n}(\vec{\lambda}) 
- t_*\omega_n(\vec{\lambda})]+\theta ~\mathrm{mod}\,2\pi, \label{eq:phases}
\end{eqnarray}
can be satisfied closely enough given $\varepsilon$. In such a case, the evolved state $\ket{\psi(t_*)}=e^{-it_* H(\vec{\lambda})}\vert i\rangle$ and the target  $\vert f \rangle$ have nearly the same expansion coefficients in the eigenbasis of $H(\vec{\lambda})$.

The system of Eqs.~\eqref{eq:distribution} shows that the probability distributions over the eigenstates must agree, i.e., the populations in the eigenbasis of $H(\vec{\lambda})$ have to be identical for the initial state $\vert i \rangle$ and the target state $\vert f \rangle$. After all, eigenstates do not evolve under their respective Hamiltonian up to a phase.
In addition Eq.~\eqref{eq:phases} states that for two states to coincide, the spectrum $\{\omega_n(\vec{\lambda})\}_{n=1}^d$ of $H(\vec{\lambda})$ must be tuned so that it matches the phase differences between the initial and final state at time $t_*$.
Note that time only appears in Eqs.~\eqref{eq:phases}, while Eqs.~\eqref{eq:distribution} depend only on initial and final state.

Therefore the question of reachability is split into two tasks: solving  $d$ nonlinear equations involving (i) the probabilities (Eqs. \eqref{eq:distribution}), and (ii) the phases (Eqs. \eqref{eq:phases}). Both involve $K$ variables $\{\lambda_\ell\}_{\ell=1}^K$; the equation for phases additionally relies on time $t_*$ and global phase $\theta$.

In this work, our goal is not to solve this system in full generality, but rather to understand what restricting the couplings to be constant in time costs, relative to time-dependent control over the same generators. A more tractable way to probe this cost is to focus on \emph{unreachability},
that is, to characterize when a given pair of states is not connected by an autonomous evolution of the form in Eq.~\eqref{eq:base_setting}. 
Equivalently, we ask under which conditions the above system of equations has no solution. 
While this may appear, at first sight, to be a problem of comparable difficulty, it turns out that detecting unreachability is substantially more accessible. 
From the viewpoint of Eqs.~\eqref{eq:distribution} and \eqref{eq:phases}, a state $\vert f \rangle$ is not autonomously reachable from $\vert i \rangle$ whenever at least one of the following \textit{obstructions} occurs: 
either the phase-matching conditions in Eq.~\eqref{eq:phases} admit no solution for any choice of control parameters $\vec{\lambda}$, or the eigenstate populations fail to satisfy the constraints in Eq.~\eqref{eq:distribution}. 

Here our special focus will be on the obstructions in Eqs.~\eqref{eq:distribution}, to match the populations of initial and final state in the eigenbasis of the $\vec{\lambda}$ controlled Hamiltonian $H(\vec{\lambda})$.

\subsection{ Sampling protocol for unreachability of typical states}\label{sec:unreach_typical}

In a specific system, the initial state and the available control terms are fixed by the platform, while the target state could be variable depending on particular needs, e.g., the unknown ground state of a problem Hamiltonian~\cite{tiwari_time_efficient_2025}.
Our aim is a statement about autonomous control, independent of the implementation details. With this in mind we fix the initial state but declare the control set itself random: the Hamiltonian terms $\{h_\ell\}_{\ell=1}^K$ are sampled from a fixed pool, and not optimized for any particular target. This way we knowingly disregard the finer structure of the Hamiltonian families, but any results we obtain are valid for \emph{typical} target states and \emph{typical} Hamiltonian families with a particular structure.

The basic procedure of the protocol is to repeat the following points:

\begin{enumerate}[label=(\alph*), ref=(\alph*)]
    \item Sample $K$ operators from a given pool, obtaining $\{h_\ell\}_{\ell=1}^K$. Check whether the sampled control set fulfills auxiliary criteria, ensuring that -- for a fair comparison -- such a family could reach any state, provided full time-dependent (non-autonomous) control (see Sec. \ref{sec:dla} below). \label{prot:hamsampl}
    \item Set the initial state as $\ket{i}=\ket{0}$ for the qudit pool, $\ket{i}=\ket{0}^{\otimes n}$ for the Pauli pool. \label{prot:initst}
    \item Sample the target state at random as $\ket{f}=U_{\text{Haar}}\ket{i}$, where $U_{\text{Haar}}$ denotes a Haar random unitary. States defined this way are uniformly spread across the Hilbert space. \label{prot:targetst}
    \item Decide whether the final state is autonomously reachable (or provably unreachable) from the initial one given sampled operators, using one of the criteria developed in the later sections. \label{prot:check}
\end{enumerate}

The selection in the step \ref{prot:hamsampl} is relatively easily checked but serves an important purpose. Without it, the sampled control set may simply not be expressive enough: some $\{h_\ell\}_{\ell=1}^K$ cannot reach the target even if their couplings were allowed to vary in time. Having such cases in the calculations would skew the results; to obtain a clear view on the effect of the restriction to autonomous evolution, we only allow the cases where the control set could reach any state given time control.

For fixed $K$, the above procedure is repeated -- new control sets of $K$ Hamiltonian terms are sampled, alongside with states, reachability of which is verified -- until a satisfactory number of samples $N$ is taken. Out of those, $N_{\text{unreach}}$ were found to be unreachable; the \emph{probability of unreachability} reported in the plots is
\begin{equation} \label{eq:punreach}
    p_{\text{unreach}} = \frac{N_{\text{unreach}}}{N},
\end{equation}
shown with $95\%$ confidence intervals. Note that this number estimates the probability that a \emph{random state} is unreachable from fixed $\ket{i}$ using \emph{randomly sampled control sets}, subject to constraints, and does not separate different sources of unreachability. For instance, $p_{\text{unreach}}=\frac12$ could mean that half of the \textit{state space} is unreachable for any choice of control terms, or that half of the sampled \textit{control terms} lead to complete reachability, and the other half to complete unreachability. In practice, $p_{\text{unreach}}$ is often very close to either 0 or 1, where this ambiguity does not arise; since the averaged quantity is easier to interpret and numerically implement, we use it throughout.

\subsection{Dynamical Lie Algebra Criterion for Non-Autonomous Reachability} \label{sec:dla}

Step~\ref{prot:hamsampl} of the protocol requires a test of whether a control set of Hamiltonian terms $\{ h_1,\ldots h_K\}$ is expressive enough to reach every state when its couplings are allowed to vary in time. 
For this \emph{non-autonomous} control problem the answer is provided by control systems theory~\cite{JurdjevicSussmann1972,DAlessandro2021}, which we summarize here.

Time dependence enlarges the reachable set because switching between generators produces their commutators. 
To get a feeling for the underlying mechanism, consider four short pulses of duration $\delta t$, generated by $h_a$, $h_b$, $-h_a$, and $-h_b$ in turn, which gives
\begin{equation*}
e^{-i h_a \delta t}\, e^{-i h_b \delta t}\, e^{i h_a \delta t}\, e^{i h_b \delta t}
= e^{-[h_a, h_b]\,\delta t^2 + O(\delta t^3)} , 
\label{eq:groupcommutator}
\end{equation*}
where we used the Baker-Campbell-Hausdorff formula.
This time dependent pulse sequence effectively amounts to an evolution under the effective Hermitian generator $-i[h_a,h_b]$ for a time $\delta t^2$, and
nesting such sequences generates nested commutators.
The \emph{dynamical Lie algebra} (DLA) of a control set is the set of all real linear combinations of the generators and all their nested commutators
\begin{equation*}
\begin{aligned}
\mathrm{DLA}(\{h_{\ell}\}_{\ell=1}^K) = \operatorname{span}_{\mathbb{R}}\bigl\{\, & i h_\ell,\ [i h_a, i h_b],\\
& [i h_a,[i h_b, i h_c]],\ \ldots \bigr\} .
\end{aligned}
\end{equation*}
With time-dependent couplings, one can implement exactly the unitaries generated by the DLA, and no others~\cite{JurdjevicSussmann1972,DAlessandro2021}.
If $\mathrm{DLA}(\{h_\ell\}_{\ell=1}^{K}) = \su(d)$, every unitary, and hence every state, can be reached, and we call the control set 
\textit{DLA complete}.
In the protocol we describe in Sec.~\ref{sec:unreach_typical}, every uniformly sampled set of control Hamiltonians is checked for having full non-autonomous control capabilities before it is scrutinized for autonomous reachability.

\section{Variance Criterion for Certified Unreachability}
\label{sec:variance}
\subsection{Construction of Variance Criterion} 
\label{sec:variance_crit}
In this section we construct an efficiently computable criterion for unreachability.
Of the two obstructions identified at the end of Sec.~\ref{sec:setup} (unsatisfiable phase-matching conditions, Eq.~\eqref{eq:phases}, and mismatched eigenstate populations, Eq.~\eqref{eq:distribution}) we exploit the population obstruction, and derive a condition for unreachability based on the energy distributions $\{p_{i,n}(\vec{\lambda})\}$ and $\{p_{f,n}(\vec{\lambda})\}$ of the initial and the target state.

These distributions coincide if and only if all their moments, defined by
\begin{eqnarray}
\mathbb{E}_\alpha \omega^m & \equiv & \sum_n (\omega_n)^m\, p_{\alpha,n}(\vec{\lambda})  = \langle \alpha \vert (H(\vec{\lambda}))^m  \vert \alpha \rangle, \nonumber
\end{eqnarray}
coincide for all $m \ge 1$. 
Conversely, the distributions differ -- and thus $\vert f \rangle$ is unreachable from $\vert i\rangle$ -- whenever at least one of these moments is different for all non-zero choices of $\vec{\lambda}$. 
Checking all moments is in general intractable, so we focus on the lowest non-trivial ones and use them as witnesses of unreachability. 

In particular, we consider the first two moments,
\begin{eqnarray}
\mathbb{E}_i \omega(\vec{\lambda}) & = & \mathbb{E}_f \omega(\vec{\lambda}) , \label{eq:1st_moment}\\
\mathbb{E}_i \omega^2(\vec{\lambda}) & = & \mathbb{E}_f \omega^2(\vec{\lambda}), \label{eq:2nd_moment}
\end{eqnarray}
and study under which circumstances they do not coincide. 
We start by rewriting the first moments as
\begin{eqnarray} \label{eq:1st_moment_expand}
\mathbb{E}_\alpha \omega(\vec{\lambda}) 
& = & \sum_{n=1}^d\omega_n(\vec{\lambda})\, p_{\alpha,n}(\vec{\lambda})
=  \sum_{\ell=1}^K \lambda_\ell \underbrace{\langle \alpha \vert h_\ell \vert \alpha \rangle}_{\equiv (\vec{v}_\alpha)_\ell}\; \nonumber \\
& = & \vec{v}_\alpha^\top \cdot \vec{\lambda} ,
\end{eqnarray}
so that the condition in Eq.~\eqref{eq:1st_moment} for some $\vec{\lambda}_*$ becomes
\begin{equation}
(\vec{v}_f - \vec{v}_i)^\top \vec{\lambda}_* = 0. \label{eq:1st_linform}
\end{equation}
Thus, any $\vec{\lambda}_*$ that is orthogonal to $\vec{v}_f - \vec{v}_i$ is a candidate solution as far as the first moments are concerned. 
Geometrically, Eq.~\eqref{eq:1st_linform} defines a hyperplane in control space on which the mean energies of $\vert i \rangle$ and $\vert f \rangle$ coincide. 
Let us denote this hyperplane -- a linear subspace in the space of parameters -- as $V_*$ for later use.

The criterion for the second moment in Eq.~\eqref{eq:2nd_moment} becomes
\begin{eqnarray}
\mathbb{E}_\alpha \omega^2(\vec{\lambda}) 
& = & \langle \alpha \vert H^2(\vec{\lambda}) \vert \alpha \rangle  = \vec{\lambda}^\top\cdot  Q_\alpha \cdot \vec{\lambda},
\end{eqnarray}
where, in analogy to Eq.~\eqref{eq:1st_moment_expand}, we expand all terms in $\langle \alpha \vert H^2(\vec{\lambda}) \vert \alpha \rangle$ and identify the matrix
\begin{equation}
(Q_\alpha)_{ab} = \frac{1}{2}\langle \alpha \vert  h_a h_b+h_b h_a \vert \alpha \rangle,
\end{equation}
which is symmetric, $Q_\alpha^\top= Q_\alpha$: since $\vec{\lambda}$ is real, only the symmetrized part of $\langle \alpha \vert h_a h_b \vert \alpha \rangle$ contributes to the quadratic form $\vec{\lambda}^\top Q_\alpha \vec{\lambda}$, which is why $Q_\alpha$ is defined through the anticommutator.
From Eq.~\eqref{eq:2nd_moment} and the condition for the equality of the first moments in the form of Eq.~\eqref{eq:1st_linform}, the joint criterion for equality of the first and second moments can be written in the compact quadratic form
\begin{equation}\label{eq:diffQ}
\vec{\lambda}_*^\top \cdot \bigl(\underbrace{Q_f - Q_i}_{\equiv \Delta Q}\bigr)  \cdot \vec{\lambda}_* = 0,
\end{equation}
where the subscript in $\vec \lambda_*$ denotes that we consider only those vectors which belong to $V_*$, that is, those for which Eq. \eqref{eq:1st_linform} holds.
Unreachability, in this setting witnessed by the failure of the second moments to coincide, can therefore be tested by checking whether the operator $\Delta Q$ restricted to the subspace $V_*$ is strictly definite:
\begin{equation}
    \Delta Q\vert_{V_*} > 0
    \quad\text{or}\quad
    \Delta Q\vert_{V_*} < 0,
\end{equation}
that is, whether all eigenvalues of the restricted operator are strictly positive, or all strictly negative. This is because if a quadratic form is strictly definite -- positive or negative -- it can only be zero for $\vec\lambda=(0,\ldots,0)$.

\begin{proposition}[Variance certificate]\label{prop:variance}
If $\Delta Q\vert_{V_*} > 0$ or $\Delta Q\vert_{V_*} < 0$, then there is no non-zero
$\vec{\lambda}$ for which the first and second moments of the two energy
distributions simultaneously agree, and $\vert f \rangle$ is not
autonomously reachable from $\vert i \rangle$.\footnote{Numerically we
test definiteness of the quadratic form
$\Delta Q + \gamma\, (\vec{v}_f - \vec{v}_i)(\vec{v}_f - \vec{v}_i)^\top$ with $\gamma$ being either a large positive or negative number. Definiteness of either of the cases certifies the stronger condition that $\Delta Q\vert_{V_*} > 0$, or $\Delta Q\vert_{V_*} < 0$, respectively; the additional term influences the direction
perpendicular to the subspace $V_*$.}
\end{proposition}
\begin{proof}
At fixed $\vec{\lambda}$, the $m$-th energy moment of a state is $\sum_n \omega_n^m\, p_n$. For $m=0,\ldots,d-1$ these are $d$ linear equations for the $d$ populations $p_n$, invertible if all $\omega_n$ differ (Vandermonde matrix), so matching the populations [Eq.~\eqref{eq:distribution}] is equivalent to matching all moments, and in particular requires equal first and second moments. Equal first moments confine $\vec{\lambda}$ to $V_*$, where the definite $\Delta Q$ makes the second moments differ for every $\vec{\lambda}\neq\vec 0$. Hence the populations never match, and $\vert f\rangle$ is unreachable. This also holds in the approximate sense of Definition~\ref{def:reach}: evolved states share the moments of $\vert i\rangle$, states close to $\vert f\rangle$ have moments close to those of $\vert f\rangle$, and for unit $\vec{\lambda}$ (which suffices, since only $t\vec{\lambda}$ matters) the moment mismatch is continuous and nonzero, hence bounded from below by a positive constant. Evolved states therefore cannot come arbitrarily close to $\vert f\rangle$.
\end{proof}

While this criterion is easily computable and readily applicable once a pair of states and control Hamiltonians are specified, it is a sufficient but not necessary test of unreachability: the two distributions may agree in their two lowest moments for some $\vec{\lambda}$ while an obstruction persists in a higher moment, to which the criterion is oblivious, so it underestimates the number of unreachable states.
In fact, it certifies nothing once the control set is large enough.

\begin{proposition}[Cutoff of the variance certificate]\label{prop:cutoff}
If $K > 2d-3$, then for any Hermitian $\{h_\ell\}_{\ell=1}^K$ and any states $\vert i\rangle$, $\vert f\rangle$, $\Delta Q\vert_{V_*}$ is neither positive nor negative definite.
\end{proposition}

The proof is given in Appendix~\ref{app:cutoffproof}.
Physically, autonomous evolution changes neither the mean nor the variance of the energy, so the criterion certifies $\vert f\rangle$ as unreachable only if no choice of couplings gives the two states equal mean energies and equal energy variances.
For $K> 2d-3$ such a choice always exists, so the criterion can no longer certify any state. 

\begin{figure}[t]
\centering
\includegraphics[width=\columnwidth]{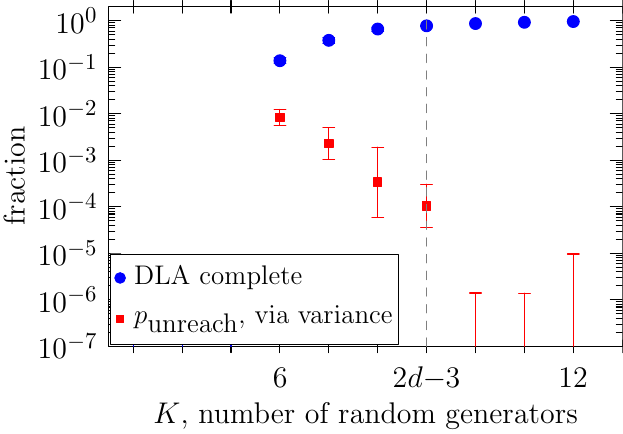}
\caption{{Certified unreachability among fully controllable draws ($d = 6$, qudit pool).} Blue circles: the fraction of Hamiltonian families with $K$ generators, sampled from the overall pool, that a full DLA analysis shows to be fully controllable when the control weights are allowed to vary in time. Red squares: the fraction of uniformly sampled states provably unreachable, by the variance criterion, under the Hamiltonian families of the previous step. Both datasets start at $K=d=6$, since no families with full DLA were found below this. 95\% confidence intervals are shown for both datasets. Beyond $K=2d-3$ (dashed line)
no state is certified unreachable, as required by Proposition~\ref{prop:cutoff}. 
}%
\label{fig:cert}
\end{figure}

\subsection{Typical Unreachable States from Variance Criterion}
The variance criterion developed in the previous section~\ref{sec:variance_crit} is an easily computable measure (not requiring optimization) to decide whether $\vert f
\rangle$ cannot be reached from $\vert i\rangle$ given a set of $K$ Hamiltonian control terms.

Let us now combine the variance criterion with the protocol outlined in Sec.~\ref{sec:unreach_typical} considering the \textit{qudit pool} with dimension $d=6$ and a varying number of generators $K$. As the general outline of our typical unreachability protocol suggests, we \ref{prot:hamsampl} first sample control terms from the pool and decide whether they have complete dynamical Lie algebra. For the samples passing the test, we construct \ref{prot:initst} fixed initial and \ref{prot:targetst} random target states, and \ref{prot:check} decide -- using the Prop. \ref{prop:variance} -- whether they are unreachable from the fixed initial state. 

The results are summarized in Figure~\ref{fig:cert} and are indicative of a typical structure for different dimensions. The first samples of complete DLA appear at $K_*=d=6$.
At this lowest $K$, we observe the most unreachable states. This tells us that the criterion is not vacuous: it can indeed certify unreachability in the nontrivial case, where, if the interaction weights were time-controlled, each state would be reachable. From there, the decline of the certifiably-unreachable state fraction is roughly exponential with increasing number of Hamiltonian terms $K$. The semi-exponential falloff continues until $K=2d-3$, beyond which ($K> 2d-3$) no states are empirically certified unreachable by the variance criterion. 

This cutoff is explained by Proposition~\ref{prop:cutoff}: for $K> 2d-3$ the variance criterion cannot certify any state, whatever the generators and the states. The data show that the bound is sharp, since certificates still occur at $K=2d-3$. Near the cutoff the certified fraction is already small, so many samples are needed to resolve it, which in practice limited us to observing the cutoff for up to $d= 10$.

\section{Spectral criterion for unreachability}\label{sec:spectral_crit}
\label{sec:breakdown}
\subsection{Spectral criterion through numerical search} \label{sec:numspecdef}
\begin{figure*}[t]
\centering
\includegraphics[width=.9\linewidth]{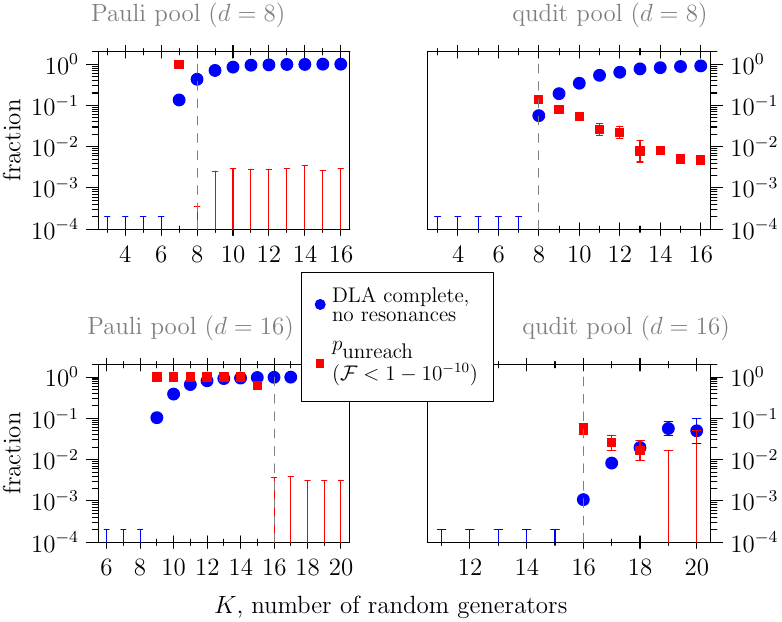}
\caption{{The unreachability transition at $d=8$ (top row) and $d=16$ (bottom row). For each number of random generators $K$, Hamiltonian and target state pairs are sampled as described in Sec. \ref{sec:unreach_typical}, so that full controllability and lack of linear dependence of eigenvalues, for both Pauli (left column) and qudit pools (right column) is assumed.} Blue circles: fraction of accepted Hamiltonian samples. Red squares: the fraction $p_{\text{unreach}}$ of uniformly sampled random target states declared unreachable by the spectral criterion as having numerically optimized fidelity $\mathcal{F}<1-10^{-10}$. Dashed line: indicates $K=d$.
By the argument presented in the text, the states found to be not unreachable are in fact reachable, within the prescribed accuracy. It can be observed that irrespective of the nature of the Hamiltonian family, first reachable states are found at $K=d$.  95\% confidence intervals are superimposed on the datapoints; due to the logarithmic scale, in the cases where either fraction of accepted Hamiltonian or probability of unreachability is zero, only the upper bound of the confidence interval is visible.
}

\label{fig:transition}
\end{figure*}
The variance criterion to determine unreachability between a pair of states efficiently detects an obstruction to find a solution to Eqs.~\eqref{eq:distribution} and does so without optimization. 
On the flip side it misses a large fraction of states which are unreachable but are not recognised by it since it only takes into account the two lowest moments
of the energy distributions.

In this section we therefore test the population-matching condition of Eq.~\eqref{eq:distribution} directly, via the fidelity
\begin{equation}
    \mathcal{F}(\vec{\lambda}) = \sum_{n} \sqrt{p_{f,n}(\vec{\lambda})}\,\sqrt{p_{i,n}(\vec{\lambda})},
    \label{eq:specfid}
\end{equation}
known in probability theory as the Bhattacharyya coefficient. Under certain assumptions, it is the tight upper bound to the fidelity between states on the orbit and the target state. Thus, the target is unreachable whenever
\begin{equation}
    \sup_{\vec\lambda} \mathcal{F}(\vec{\lambda}) < 1,
\end{equation}
 this line of thought was first explored in Ref.~\cite{schirmer_design_2018}.
This value depends on $\vec{\lambda}$ only through the eigenbasis of $H(\vec{\lambda})$: it is invariant under the rescaling $\vec{\lambda} \to c\vec{\lambda}$, and the evolution time does not enter it, so the search effectively runs over the direction of $\vec{\lambda}$ alone.

Strictly speaking, testing this criterion requires finding the maximum over the entire parameter space $\mathbb{R}^K$.
In practice, we declare unreachability when exhaustive gradient-based maximization with multiple restarts from random initial parameter values fails to bring the numerical optimum $\mathcal{F}_*$ close to unity within a fixed tolerance $\varepsilon$; we write $p_{\mathrm{unreach}}(K)$ for the fraction of random instances with $K$ generators for which the target is declared unreachable.

In the limit $\varepsilon \to 0$ this probes reachability in the strict sense of vanishing orbit distance (Def. \ref{def:reach}). Empirically, we see that for states explicitly constructed as reachable, numerically perfect optima are found if sufficiently many initial parameter distributions are used to start the optimization. This strategy defines the computational budget used to determine unreachability (see Appendix~\ref{app:optimdetails}). It is however always possible that the numerical optimization fails to find the global optimum, so $p_{\mathrm{unreach}}$ formally is just an upper bound. %

In the following, we first motivate the mathematical validity of the numerical procedure, predicting some of the resulting properties, before showing the numerical results in Sec.~\ref{sec:numericalspectral}.

\subsection{Rational independence of energies and torus fillings}

In some cases, the value of $\mathcal{F}$ at $\vec\lambda$ corresponds to the supremum\footnote{Strictly speaking, \emph{maximum} often does not exist, just as the function $\sin(x)+\sin(\sqrt{2} x)$ never reaches 2, but for any $\varepsilon>0$, there exists a $x_*$ for which the value of the function is $2-\varepsilon$; in this case, $2$ is the supremum.} of the fidelity over the orbit, so that
\begin{equation}
\label{eq:fidsup}
\mathcal{F}(\vec\lambda) = \sup_t \left\vert \langle f \vert e^{-iH(\vec\lambda) t} \vert i \rangle \right\vert.
\end{equation}
Whether the above identity holds or not is controlled by the \emph{rational independence} of the eigenvalues of the Hamiltonian $H(\vec\lambda)$. Intuitively, if such rational dependencies exist, some of the parts of the evolved vector $\exp(-iH(\vec \lambda) t)\ket{i}$ are resonant to each other, and the orbit does not explore the entire space of possibilities. If, on the contrary, no such dependencies exist, the orbit covers the space of possibilities densely, and any state $\ket{f}$ with $p_{f,n} = p_{i,n}$ lies infinitely close to an orbit. 

Let us formalize the above notion.
\begin{definition}
\label{def:ratdep}
The eigenenergies $\vec\omega$ of the Hamiltonian $H$ are said to be rationally independent if and only if 
\begin{equation}
\sum_{n=1}^d \omega_n m_n \neq 0,
\label{eq:ratdep}
\end{equation}
for any nonzero rational vector $\vec m\in\mathbb{Q}^d$ orthogonal to $(1,1,\ldots, 1)$.
\end{definition}
Each rational vector $\vec m$ violating Eq. \eqref{eq:ratdep} constrains the orbit, in a sense spelled out below. The orthogonality constraint then plays an important role of omitting the unphysical global phase from the considerations: traceless Hamiltonians possess a trivial eigenvalue dependence $\sum_{n} \omega_n = \operatorname{Tr} H =0$, but this effect is not observable.

To study the effects of rational dependence, let us introduce the set of evolved states 
\begin{equation}
\begin{aligned}
\mathcal{E}_{H,\vert i\rangle} = \Big\{e^{i\alpha}\vert \psi(t)\rangle &=  e^{i\alpha}e^{-iHt }\vert i\rangle\\& : t\ge 0, \alpha \in \mathbb{R} \Big\}
\end{aligned}
\end{equation}
One might expect the set of time-evolved states $\mathcal{E}_{H,\vert i\rangle}$ to be of low dimension, since the time $t$ and overall phase $\alpha$ are the only free parameters. This is true, but the set of states infinitely close to it has higher dimensionality, and the reason is the closure in Definition \ref{def:reach}. Such states may not need to lie on any single orbit point, but can be approximated arbitrarily close by the evolved states. This happens exactly if there exist no rational dependencies in the spectrum as specified by Definition \ref{def:ratdep} -- see Fig. \ref{fig:concept} for a schematic picture of this phenomenon.
To see this, we start by rewriting the time-evolved state of Eq.~\eqref{eq:polar} as
\begin{equation}
    \vert \psi(t)\rangle  = \sum_{n=1}^d {\sqrt{p_{i,n}}\vert n\rangle} e^{-i\varphi_n(t)}
\end{equation}
with the time-dependent vector of phases $\vec{\varphi}(t) = \vec{\omega}t + \vec{\varphi}_0$, which traces a curve on the torus of phases $T_d = [0,2\pi)^d$.
The set $\mathcal{E}_{H,\ket{i}}$ is therefore a subset of the more general set of torus states, one for every point on the torus:
\begin{equation}
    \mathcal{T}_{H,\ket{i}} = \left\{\vert \psi(\vec{\varphi})\rangle = \sum_{n=1}^d \sqrt{p_n}\vert n\rangle e^{-i\varphi_n}: \vec{\varphi}\in T_d \right\}
    \label{eq:torusstates}
\end{equation}
We see that the time evolution is just a special case of this more general set of wave functions on the torus,
\begin{equation}
    \vert \psi(t)\rangle = \vert \psi (\vec{\varphi}(t))\rangle.
\end{equation}

Let us now study the curve $\vec{\varphi}(t)$ through $\vec{\varphi}_0$, together with all the points it comes arbitrarily close to, by considering the \emph{closure} of~the~curve~$\{\vec\varphi(t)\}$:\footnote{The closure of any set $S$ is denoted by $\overline S$.}
\begin{equation}
    \mathcal{C}_{\vec{\varphi}_0} = \overline{ {\{\vec{\varphi}(t) =\vec{\omega}t + \vec{\varphi}_0: t\in \mathbb{R}\}}}.
\end{equation}

The question whether the above set is equal to the torus $T_d$ has been extensively studied in the number theory and classical mechanics literature. The criterion, adapted from Theorem 444 of \cite{hardywright}, can be stated as the following:
\begin{proposition}
    
\label{thm:torusstates}
    If the eigenvalues $\vec\omega$ of $H$ are rationally independent, the closure of the set of evolved states is the set of torus states, $\overline{\mathcal{E}_{H,\ket{i}}} = \mathcal{T}_{H,\ket{i}}$.
\end{proposition}

The above result, applied to the numerically found optimum of $\mathcal{F}$ at $\vec\lambda_*$, leads to the following observation.

\begin{proposition}[Sufficiency of population matching]\label{thm:sufficiency}
Let $\vec{\lambda}_*$ be such that the spectrum $\vec{\omega}(\vec{\lambda}_*)$
is rationally independent. Suppose further that the populations match,
$p_{i,n}(\vec{\lambda_*}) = p_{f,n}(\vec{\lambda_*})$ for all $n$. Then
$\vert f \rangle$ is autonomously reachable from $\vert i \rangle$ in the
sense of Definition~\ref{def:reach}: the orbit of $\vert i \rangle$ under
$H(\vec{\lambda_*})$ passes arbitrarily close to $\vert f \rangle$, up to a
global phase.
\end{proposition}

Rational independence thus converts matched populations into full reachability. However, the question whether a given numerical optimum actually has a rationally independent spectrum cannot be decided in floating-point arithmetic. 
Appendix~\ref{app:resonance} gives the test we use instead: if no \emph{real} linear relation among the eigenvalues persists on an open region of parameter space (a condition that is numerically checkable), then rational relations are confined to a closed set with empty interior, so arbitrarily close to any optimum there is a Hamiltonian with a rationally independent spectrum.

In our computations the populations $(p_{i,n})$ and $(p_{f,n})$ are matched only to a finite fidelity $\mathcal{F}\ge 1-\varepsilon$. Notice that -- if the prerequisites of the above reasoning are met -- by optimizing $\mathcal F$ over the weight vector $\vec\lambda$ we implicitly optimize over the time $t$ as well, in the sense of finding a supremum in Eq. \eqref{eq:fidsup}. Finding local maxima close to the supremum is a difficult problem: the phases $\omega_n t$ must line up with the required phase differences modulo $2\pi$ for all $n$ at once, the times at which this happens can grow rapidly with $d$, and no efficient method for finding them is known to us.\footnote{Finding the time $t_*$ for which the fidelity is close to the supremum is related to the problem of solving simultaneous Diophantine approximations, a problem whose natural formulations are NP-hard~\cite{lagarias_sda_1985}.}
This requires no further optimization of the Hamiltonian controls, only enough time for the torus-filling time evolution to come close enough to the desired point.

\subsection{Theory of the reachable set dimension}\label{sec:theory}
In the previous subsection we observed that despite a single orbit being essentially a one-dimensional curve\footnote{If the unphysical overall phase is not taken into account.}, the set of reachable states can have a higher dimension under certain circumstances, as the orbit covers an entire $d$-dimensional torus densely.  This reasoning was presented for a fixed Hamiltonian $H$. However, if the $H$ is parameterized as $H(\vec \lambda)$, the set of reachable states has generically higher dimensionality. Here we present the theory of the dimensionality of this set. The main result, reasoned below is the following:
\begin{proposition}\label{prop:dimension}
    Consider the fixed control set $\{h_\ell\}_{\ell=1}^K$ and initial state $\ket{i}$. The set union of reachable sets for all choices of $\vec\lambda$,
    \begin{equation}
        \bigcup_{\vec\lambda} \overline{\mathcal{E}_{H(\vec\lambda),\ket{i}}},
    \end{equation}
    has dimension at most $K+d-2$.  
    
\end{proposition}
Thus, the reachable set could start having full dimensionality at $\Kc = d$ generators. Our numerical results (see Sec. \ref{sec:numericalspectral}) imply that this bound is attained: a state $\ket{f}$ sampled uniformly from all pure states \emph{will have nonzero probability of being reachable} if sufficiently many generators are available. The reasoning behind this proposition is presented below.

A physical state (a ray in $\mathcal{H}_d$) is described by $2d-2$ real parameters: a $d$-dimensional complex vector has $2d$ real components; normalization removes one, and one more is removed because two vectors differing only by a global phase factor $e^{i\theta}$ describe the same physical state.
An autonomous evolution can therefore produce arbitrary states only if it supplies at least $2d-2$ real parameters.

Let us start by splitting the vector of coupling constants into its length times a unit vector, $\vec{\lambda} = \vert \vec{\lambda}\vert \cdot \vec{e}_{\vec{\lambda}}$.
The length $\vert \vec{\lambda}\vert$ will simply renormalize the evolution time, $t \rightarrow \tilde{t} = \vert \vec{\lambda}\vert t$.
In the following we will, without loss of generality, assume that $\vec{\lambda}$ is a unit vector.
As with a normalized state in a vector space, the normalization constraint removes one free parameter, so we need $K-1$ real parameters to fully characterize a unit vector\footnote{These real parameters are the generalized angles $\vec{\lambda} = \vec{\lambda}(\theta_1,\ldots,\theta_{K-1})$, where we have the azimuthal angles $\theta_1,\ldots,\theta_{K-2}$ and one polar angle $\theta_{K-1}$.}.
This shows that the autonomous control Hamiltonian, with $K$ Hamiltonian control terms, is fully characterized by $K-1$ real parameters,
\begin{equation}
    H = H(\theta_1,\ldots,\theta_{K-1}),
\end{equation}
since only the direction of $\vec{\lambda}$, not its length, matters.

For each of the choices of the weights $\vec \lambda$, the reachable states form a torus -- see Proposition \ref{thm:torusstates}. The torus is parametrized by $d$ real phases, one of which is the unphysical global phase: the physically distinct wave functions on the torus form a $(d-1)$-parameter family of states.
We thus conclude that, given (generically) rational independence of the spectrum of $H(\theta_1,\ldots,\theta_{K-1})$, the closure of the set of time-evolved states supplies $d-1$ physical parameters at fixed couplings.

Altogether, then, the closure of the set of states obtained by autonomous time evolution is parametrized by the $K-1$ coupling directions $(\theta_1,\ldots ,\theta_{K-1})$ plus the $d-1$ physical torus phases; since the map $(\vec{\theta},\vec{\varphi}) \mapsto \vert\psi(\vec{\varphi})\rangle$ is well behaved away from eigenvalue crossings\footnote{Away from eigenvalues' crossings, a small change in parameters $\vec\theta$ and $\vec\varphi$ produces a small change of the state (at most proportional to the parameter distance) -- it is locally Lipschitz continuous. Such a map can not "inflate" its domain like space filling curves do. Strict mathematical treatment of crossings is difficult, however they occupy a low-dimensional portion of the $\vec\theta$ domain, and intuitively, for each type of crossing, a similar reasoning applies, with some of the phases $\vec\varphi$ being dependent on the rest.}, the image of this parametrization cannot have dimension larger than its domain, $K+d-2$.

For the resulting set of states to have nonzero volume in the set of all pure states, described by $2d-2$ parameters, we thus need
\begin{equation}
(K-1) \, + \, (d-1) \ge 2d-2.
\end{equation}
The critical number of generators $K_c$ is obtained at equality in this expression, giving
\begin{equation}
    K_c = d,
\label{eq:Kc}
\end{equation}
which is consistent with the numerical findings, presented in the next subsection.

\subsection{Typical unreachable states} \label{sec:numericalspectral}

Unlike the variance criterion, the spectral criterion
of Sec.~\ref{sec:numspecdef} relies on numerical
optimization, so we carefully scrutinize its validity.
With insufficient optimization we would still obtain
data, namely a best fidelity $\mathcal F_*$ for each
target $\ket{f}$ and control set
$\{h_\ell\}_{\ell=1}^K$, but without assurance that
they reflect the structure of the reachable set,
such as its dimension. Insufficient optimization can only misclassify a state as unreachable, thus overall increasing $p_{\mathrm{unreach}}$ [Eq.~\eqref{eq:punreach}] and never
lowering it. We verify that the optimization is
sufficient in low dimensions, by checking that
states constructed to be reachable are found as
such (see App. \ref{app:optimdetails}).

The spectral criterion is applied within the protocol of Sec.~\ref{sec:unreach_typical}. For each pool (Pauli strings and the qudit pool of Sec.~\ref{subsec:ham_families}) we fix the dimension $d$ and the number of generators $K$; then,
\ref{prot:hamsampl} sample $K$ control terms until we
find a set with full dynamical Lie algebra and with
certifiably no rational eigenvalue dependence (generically; see App.~\ref{app:resonance}).
Then \ref{prot:initst} the initial and
\ref{prot:targetst} random target states are
constructed, and \ref{prot:check} the fidelity of Eq.~\eqref{eq:specfid} is maximized over $\vec\lambda$; the target is declared unreachable
if the optimum $\mathcal F_*$ falls below a fixed
threshold. 

In Figure \ref{fig:transition} the results for $d=8$ are presented, with the threshold of $\mathcal{F}_* < 1-10^{-10}$ used to declare a state unreachable.\footnote{This threshold, while very strict, is within the numerical accuracy of our method.}

On the one hand, for the qudit pool, the first Hamiltonian samples passing the tests above are found at $K=d=8$, and immediately, most (but not all) of the states are found reachable. With increasing number of generators $K$, the probability of unreachability $p_{\text{unreach}}$ quickly decreases, but stays nonzero even for larger $K$.

On the other hand, the first passing Hamiltonians for the Pauli pool appear at $K=7=2n+1$ (Sec.~\ref{sec:comparison_DLA}). As $K<d$, Proposition~\ref{prop:dimension} implies $p_{\text{unreach}}=1$ for exact reachability,
as the reachable set has lower dimension than the ambient  state space. Numerics agrees with the mathematical reasoning; $1-p_{\text{unreach}}\approx 1\%$.\footnote{Stricter threshold removes the small discrepancy, and we believe it is caused by the fact that the reachable set is close to large parts of the state space -- see Figure \ref{fig:fourqubitdata} of Appendix \ref{app:optimdetails} for the analysis of $d=16$-dimensional four qubit case, where this effect is prominent.} This is not a failure of the optimization procedure: at this threshold, overwhelming majority of states explicitly constructed as $\exp(-i\sum_{\ell=1}^K \lambda_\ell h_\ell)\ket{i}$ with random weights $\vec\lambda$ are found reachable. A dramatic change appears at $K=d$, where all states are found reachable.\footnote{This pattern for Pauli pool is repeated for $d=16$, 4 qubits -- see Appendix \ref{app:optimdetails}.}

The data gathered here suggests that the bounds developed in Sec. \ref{sec:theory} are tight, since in both analyzed Hamiltonian pools, full dimensional reachable sets appear at $K=d$. {For less strict fidelity thresholds the picture is different. For smallest control set size $K$ allowing \emph{time-dependent} control, every sampled state has $\mathcal{F}_*>0.99$ -- so it lies close (in fidelity) to the strictly reachable set. We call such states \emph{approximately reachable}. This is true for both pools at $d=8$ shown in Figure \ref{fig:transition} and higher dimensions: four-qubit Pauli pool (Fig. \ref{fig:fourqubitdata}) and for the limited number of instances\footnote{Total of 170 random states per pool.} we could optimize at $d=32$ (both pools; for Paulis it corresponds to five qubits). The found $\mathcal{F}_*$, while being just local optima and thus just lower bounds to true global maximum, are high enough for e.g. state preparation. }

{This behavior does not contradict the dimension count of Sec. \ref{sec:theory}: the states within the given fidelity threshold to a lower-dimensional reachable set still form a set of nonzero volume. For more details on how the volume depends on the threshold, see Appendix \ref{app:optimdetails}.}%

\subsection{Mean time of reachability}\label{sec:meantime} 

We consider a fixed initial and final state whose populations $\{p_n\}$ coincide by construction, with sufficient control degrees of freedom to enforce this matching. We assume the spectrum $\{\omega_n\}$ is rationally independent so that the phase flow $\vec{\varphi}^\prime(t) = \vec{\nu}t +\vec{\varphi}_i^\prime$ is ergodic on the $(d-1)$-dimensional torus. An arbitrary global phase, identified with the $d$-th component, has been subtracted: $\nu_n = \omega_n-\omega_d$ and $\varphi^\prime_{\alpha,n} = \varphi_{\alpha,n}-\varphi_{\alpha,d}$ 
where $\alpha=i,f$ and therefore the dimensionality is reduced by one. 

For a given final state the question is how long the phase trajectory takes to come $\epsilon$-close to the target phase point. This is difficult for a single final state, but {a related problem} becomes tractable upon averaging over the ensemble of typical final states.  For a Haar-distributed final state $\vert f\rangle =\sum_n \sqrt{p_n}e^{-i\varphi_{f,n}}\vert n\rangle$ with fixed $p_n$, the phase vector $\vec{\varphi}^\prime_{f}$ is uniformly distributed on the torus: $\rho_{\rm Haar}(\vec{\varphi}_f^\prime) = \rho_0 = (2\pi)^{1-d}$ \cite{mehta2004random,forrester2010log}.

To simplify the geometry of the curve we periodically unfold the torus onto $\mathbb{R}^{d-1}$ modular space.  
The orbit becomes a straight line with velocity $\vec{\nu}$, and the typical target states becomes a {randomly translated grid of balls of radius $\epsilon$, one ball per cell of size $2\pi$. In this picture the trajectory comes $\epsilon$-close to a final state precisely when it scatters off an $\epsilon$-ball centred on that state. The mean reachability time is therefore the mean scattering time $t^*$.}

{In the heuristic reasoning below we replace the grid with} uniformly distributed balls of radius $\epsilon$ with the same density $\rho_0$.

In the following we present a physicist friendly derivation of the results in \cite{dumas1991ergodization,bourgain1998distribution} for estimating the mean free path $L^*$. 
The mean free path is the average distance a point particle travels through a medium of uniformly distributed $\epsilon$-balls before scattering. This is a standard problem in classical kinetic theory \cite{huang2010statistical}. In a cylindrical tube of length $L$ and cross-sectional area $A$ the number of scatterers is $N_L = \rho_0 A L$. Each scatterer presents a cross section $\sigma=V_{d-2}\epsilon^{d-2}$ where $V_d = \pi^{d/2}\,/\,\Gamma(d/2 + 1)$ is the volume of the unit ball. At the mean free path the total cross-sectional area of all scatterers in the tube equals the tube cross section: $N_L \sigma \approx A$. For shorter cylinders $L<L^*$ the particle has a finite probability of traversing the full length without collision. From this we obtain the mean free path $L^*=(\sigma\rho_0)^{-1}$.

Relating the mean free path to the trajectory velocity in the unfolded moduli space gives $L^* = \vert \vec{\nu}\vert t^*$. The typical reachability time is therefore
\begin{equation}\begin{aligned}
t^* &= \mathcal{C} \,\epsilon^{2-d} = O(e^{d(\log(d)+\log(1/\epsilon))})%
\end{aligned}
\end{equation}
where $\mathcal{C}=\frac12 \vert \vec{\nu}\vert^{-1} (2\sqrt\pi)^d \Gamma(d/2)$. The reachability time depends critically on the tolerance $\epsilon$ and grows faster than exponentially with the Hilbert-space dimension $d$, the latter through the gamma function in $\mathcal{C}$ for large $d$.
So autonomous dynamics can be very expressive but comes at the high cost of being very slow in high dimensional Hilbert spaces.

\section{Comparison with Non-autonomous Control}\label{sec:comparison_DLA}

So far, we have studied the expressivity of autonomous control and found that typical states become reachable only once the number $K$ of uniformly sampled control Hamiltonians reaches the Hilbert-space dimension, $K_c = d$. This result is agnostic to the structure of the control Hamiltonians, in particular to the Lie algebra they generate. We now compare autonomous control with non-autonomous, i.e., time-dependent, control.

As a proxy for the power of time-dependent control, we use the dynamical Lie algebra (DLA) criterion introduced in Sec.~\ref{sec:dla}. 
The DLA of a control set $\{h_1,\dots,h_K\}$ is their Lie closure, i.e., the real linear span of the generators and all their nested commutators. 
Every element $H_\mathrm{DLA} \in \mathrm{DLA}(\{ h_\ell\}_{\ell=1}^K)$ can be synthesized by pulsed switching between the $h_\ell$, via the Lie--Trotter product formula and group commutators [Eq.~\eqref{eq:groupcommutator}]. Conversely, no time-dependent protocol can leave the group $e^{-iH_\mathrm{DLA}}$. 
The DLA therefore captures exactly the effective generators accessible to time-dependent driving. If $\mathrm{DLA}(\{ h_\ell\}_{\ell=1}^K) =\mathfrak{su}(d)$, every unitary, and hence every state, is reachable, and the set of controls is called DLA complete. 
DLA completeness thus serves as our proxy for full time-dependent or non-autonomous control. 
For generators sampled uniformly from the Pauli and the qudit pool, we determine how many are needed to reach it.

We begin with the Pauli pool, consisting of $n$-qubit Pauli strings. For each $K$, we sample $K$ Pauli strings uniformly at random, check whether the resulting set is DLA complete, and repeat over many realizations to obtain the probability that a random set of $K$ generators is DLA complete. The blue dots in the left panel of Fig.~\ref{fig:transition} show this probability for $n = 3$ qubits: DLA completeness sets in at $K_{\rm Pauli} = 7$. This agrees with the theoretical minimum 
\begin{equation}
K_{\rm Pauli} = 2n+1 = 2\log_2(d) +1,
\end{equation}
which follows from the algebraic structure of Pauli strings.
This minimum was established in Ref.~\cite{Smith2025} (see also Ref.~\cite{Aguilar2024}); a derivation is sketched in App.~\ref{app:dla}.
The red squares in the same panel show the probability that a state is autonomously unreachable. This probability drops once $K$ reaches the Hilbert-space dimension, $K_c = d = 8$, in line with Eq.~\eqref{eq:Kc}. Time-dependent control thus requires fewer generators than autonomous control already for three qubits. 
For $n=4$ the same onset at $K = d = 16$ is observed (Fig.~\ref{fig:fourqubitdata}); beyond that the strict-threshold numerics become unreliable, but the scaling follows from Proposition~\ref{prop:dimension}.
Expressed in terms of the Hilbert-space dimension, $K_{\rm Pauli} = 2\log_2 d + 1$, whereas autonomous control requires $K_c = d$. For this set of generators autonomous control needs exponentially more generators than time-dependent control, confirming the common expectation that time-dependent control is far more powerful.

The qudit pool tells a different story. The right panel of Fig.~\ref{fig:transition} shows the same analysis for the qudit pool of Eq.~\eqref{eq:quditpool}. 
Theoretical considerations for how many generators are needed for 
DLA completeness set the minimal number at 
\begin{equation}
K_{\rm qudit} = d 
\end{equation}
generators (blue dots). 
This result is derived in App.~\ref{app:dla} using technical results obtained in \cite{schirmerleahy2002,Schirmer2002,Altafini2002,albertini_notions_2003}.
The onset of DLA completeness is exactly the point at which autonomous control becomes effective and typical states cease to be unreachable (red squares), so $K_{\rm qudit} = K_c$. 
Remarkably, for this pool, autonomous control is therefore on par with time-dependent control in the number of generators. The reason is that each generator of the qudit pool couples only two levels: the generators must first connect all $d$ levels, and commutators cannot compensate for this sparsity~\cite{Turinici2003,Altafini2002,Schirmer2002}. 
However, the parity for this control set comes with a caveat: autonomous control matches the expressivity of its time-dependent counterpart only at long evolution times.

\section{Conclusion}
\label{sec:conclusion}
We have studied which states can be prepared by the free evolution of a Hamiltonian whose couplings are held fixed, for control sets sampled at random from two pools of elementary interactions and for typical target states. To this end, we introduced two complementary criteria of unreachability. The variance criterion (Proposition~\ref{prop:variance}) provides a certificate built from the two lowest moments of the energy distributions of the initial and target states; this certificate requires no optimization. The spectral criterion optimizes the match between the full energy distributions; for couplings with a rationally independent spectrum -- which is the generic behavior for the studied cases -- matched distributions imply that the orbit passes arbitrarily close to the target (Proposition~\ref{thm:sufficiency}).

The variance criterion certifies unreachable targets even for control sets that are DLA complete, i.e., that reach every state with time-dependent couplings. The certified fraction decays roughly exponentially with the number of generators $K$ and vanishes identically for $K > 2d-3$ (Proposition~\ref{prop:cutoff}); our data for $d\le 10$ suggests that this bound is saturated.

The reachable set of a control set of $K$ generators has dimension at most $K+d-2$ (Proposition~\ref{prop:dimension}), so typical states can be reached exactly only with $\Kc = d$ or more generators. The spectral criterion shows that this bound is attained in both pools: reachable typical targets first appear at $K=d$. What this means in comparison with time-dependent control depends on the pool (Sec.~\ref{sec:comparison_DLA}). In the qudit pool, the smallest DLA-complete control sets also have $d$ generators, and most typical targets are autonomously reachable as soon as full controllability sets in. In the Pauli pool, full controllability requires only $2\log_2 d+1$ strings, so reaching typical states exactly with fixed couplings requires exponentially more generators than reaching every state with time-dependent ones. 
Approximate reachability, however, is much less demanding: already with the smallest DLA-complete sets of Pauli strings, essentially all sampled states are approximately reachable, so that some reachable state has fidelity to the target $\mathcal{F}_*>0.99$, and for four qubits at least half of them lie above the fidelity threshold $\mathcal{F}_*>1-10^{-3}$ (Fig.~\ref{fig:fourqubitdata}). A low-dimensional reachable set can thus pass close to most of the state space, and in this approximate sense autonomous control matches the reach of time-dependent control over the same generators. For these calculations, target states are sampled uniformly, whereas ground states of local Hamiltonians, which classically optimized static Hamiltonians are designed to prepare~\cite{tiwari_time_efficient_2025}, potentially may require smaller control sets.

This comes at the cost of long evolution times: for an evolved state to approach the target within a desired accuracy, it is required that $d-1$ relative phases align, and the time at which the alignment happens grows {faster than} exponentially in the dimension $d$ (Sec.~\ref{sec:meantime}). For $K>d$, many different couplings match the populations (Appendix~\ref{app:optimdetails}), so an open question is how short the preparation time can be made, at fixed coupling strength, by choosing among them.

\section{AI usage disclosure}
Anthropic large language models (Claude Opus versions 4 to 5.5) were used for several tasks related to this paper: \begin{enumerate*}[label=(\alph*)]
\item code completion: sensitive parts of the optimization process were written by hand, and then optimized by the language model and packaged for easier use under strict supervision; \item argument in Appendix \ref{app:resonance} for reachability: the initial intuition for local detection of linear dependence of eigenvalues was proposed by one of the authors, and the idea for a proof through symmetric polynomials was devised by the model, subsequently traced back to standard literature and reconstructed from there by the authors; \item proof of Proposition \ref{prop:cutoff} in Appendix \ref{app:cutoffproof}: initial observation of $2d-3$ cutoff and interpretation of being related to the number of parameters came from the authors, the model provided linear-algebraic argument, which was redone by the authors, \item ideas for the proofs of minimal full-DLA control sets in Appendix \ref{app:dla} and \item language review of this manuscript.
\end{enumerate*} %

\section{Acknowledgements}

KS thanks Otfried Gühne, Marcus Huber, Chau Nguyen, and Marco Wiedmann for helpful discussions. DeQHOST APVV-22-0570, QUAS VEGA 2/0164/25, Postdokgrant APD0161, Štefan Schwarz programme, Deutsche Forschungsgemeinschaft (DFG, German Research Foundation, project numbers 447948357 and 440958198), the Sino-German Center for Research Promotion (Project M-0294), the ERC (Consolidator Grant 683107/TempoQ), the German Ministry of Education and Research (Project QuKuK, BMBF Grant No. 16KIS1618K), the Stiftung der Deutschen Wirtschaft, the European Union’s Horizon 2020 research and innovation programme under the Marie Skłodowska-Curie grant agreement (No 945422) supported this work.
YM would like to thank N. Haunschmied for helpful discussions on related work. YM acknowledges funding from FWF grant ISOQUANT I6949.
PR acknowledges that this project was funded in part by the Austrian Science Fund (FWF) [ESP4563925, DOI: \href{https://doi.org/10.55776/ESP4563925}{10.55776/ESP4563925}].
TA and PR acknowledge funding from the Austrian Federal Ministry of Education, Science and Research via the Austrian Research Promotion Agency (FFG) through the flagship project FO999897481 (HPQC) and the project FO999914030 (MUSIQ) funded by the European Union—NextGenerationEU.

\bibliographystyle{apsrev4-1fixed_with_article_titles_full_names_new}
\bibliography{references_cleaner}
\appendix
\section{Resonance/symmetry finding in Hamiltonian families}
\label{app:resonance}
For the proper function of the spectral criterion, we have to ascertain that the Hamiltonians found by it do not have rationally dependent eigenenergies -- that is, with 
\begin{equation}
    H(\vec\lambda)=\sum_{\ell=1}^K \lambda_\ell h_\ell=\sum_{n=1}^d \omega_n(\vec\lambda) \ket{n_{\vec\lambda}}\bra{n_{\vec\lambda}},\end{equation}

that for the optimal $\vec\lambda$, no $\vec m \in \mathbb{Z}^d$ exist such that $\sum_{n=1}^d m_n \omega_n=0$ (with some constraints, see Def. \ref{def:ratdep}).

This task, strictly interpreted, is not feasible on several grounds. We only perform numerical calculations with floating point numbers, and strict rational dependence calculations are impossible in this regime -- the floating point numbers store only approximations to the "true" numbers, and for rational dependency checks to function, it would require perfect knowledge on the true eigenvalues as real numbers. Second, even if we had an algebraic description of the optimal $\vec\lambda$, checking whether such $\vec m\in\mathbb{Z}^d$ exists is a highly nontrivial task.

Here we present a relaxation of the rational dependency checks. Instead of checking rational dependency at the optimum, we detect \emph{real linear} dependency on a selected subset of parameters $\vec\lambda$. We prove that if no such dependency can be found (which is numerically simple), then for any $\varepsilon>0$ and any $\vec\lambda$, there exists $\vec\lambda'$ such that $\vert \vec\lambda - \vec\lambda'|<\varepsilon$ and $H(\vec\lambda')$ is rationally independent. Provided spectral criterion yields a satisfactory optimum to $H(\vec\lambda)$, this proves reachability in the sense of orbits passing infinitely close to the target state, since $H(\vec\lambda')$ can be made arbitrarily close to $H(\vec\lambda)$ and still have rational independence.

First, we require some nomenclature:
\begin{itemize}
    \item Let $\Sigma$ denote "well-behaved" subsets of the parameter space $\mathbb{R}^K$: $S\in\Sigma$ if and only if it is open, connected, and $H(\vec\lambda)$ has no eigenvalue crossing for any $\vec\lambda \in S$. Since each such open set $S$ by definition avoids crossings, their set sum  $\bigcup \Sigma$ is exactly the subset of $\mathbb{R}^K$ giving rise to no eigenvalue crossing.
    \item Let $M_{\mathbb{Z}}$ denote the set of allowed integer relation vectors:
    \begin{equation}M_{\mathbb{Z}} = \{\vec m \in \mathbb{Z}^d : \sum_{n=1}^d m_n =0, \vec m\neq \vec 0\}.\end{equation}
    By setting the sum to zero, we exclude a "trivial" relation of $\vec m =(1,1,\ldots,1)$, corresponding to $\sum_{n=1}^d\omega_n =\operatorname{Tr} H(\vec\lambda)=0.$ This relation does not influence reachability; it only constraints the undetectable global phase of the states on the orbit.
    \item We will similarly denote
    \begin{equation}
    M_{\mathbb{R}} = \{\vec m \in \mathbb{R}^d : \sum_{n=1}^d m_n =0, \vec m\neq \vec 0\}.\end{equation}
\end{itemize}

\begin{proposition}If there exists a nonempty $S\in\Sigma$ and a vector $\vec m\in M_{\mathbb{Z}}$ describing rational energy dependence in $S$:
\begin{equation} F_{\vec m}(\vec \omega)=\sum_{n=1}^d \omega_n (\vec \lambda) m_n =0,~~ \forall\vec\lambda\in S,\end{equation}
then for any other $S'\in\Sigma$, there exists a vector $\vec m'\in M_{\mathbb{Z}}$ such that 
\begin{equation}\forall \vec\lambda'\in S',~~~ \sum_{n=1}^d \omega_n(\vec\lambda') m'_n=0.\end{equation}
In other words, rational dependence of eigenvalues can not be localized: if it's found on any open subset (e.g. open ball, however small), it must be found in any other parameter region (that is free of eigenvalue crossings), perhaps with different coefficients.
\end{proposition}
\begin{proof}
    Let $F_{\vec m} (\vec\omega) = \sum_{n=1}^d \omega_n m_n$. By assumption, it vanishes on $S$ if evaluated as $F(\vec\omega(\vec \lambda))$. Then let us construct a symmetrized version: it is a product of sums like in $F_{\vec m}$, but with all permutations to the indices (permutation group $\operatorname{Sym}(d)$):
    \begin{equation}{\bf F}(\vec\omega) = \prod_{\sigma \in \operatorname{Sym}(d)} \sum_{n=1}^d \omega_n m_{\sigma(n)}.\end{equation}
    Since $\bf F$ is explicitly constructed as a product of all permutations, it is a symmetric polynomial in $\vec\omega$ (that is, ${\bf F}(\omega_1,\ldots, \omega_d) = {\bf F}(\omega_{\sigma(1)},\ldots,\omega_{\sigma(d)}$ for all permutations $\sigma$). \\
    By the fundamental theorem of symmetric polynomials, $\bf F$ admits a form
    \begin{equation}{\bf F}(\vec \omega) = \operatorname{poly}(c_1(\vec\omega),\ldots,c_d(\vec \omega)),\end{equation}
    where $\operatorname{poly}(\ldots)$ is some polynomial (irrelevant), and $c_i$ are elementary symmetric polynomials:
    \begin{equation}\begin{aligned}
    c_0(\vec\omega)&=1,\\
        c_1(\vec\omega)&=\sum_{n} \omega_n,\\
        c_2(\vec\omega)&=\sum_{n_1<n_2} \omega_{n_1} \omega_{n_2},\\\vdots\\
        c_d(\vec\omega)&=\omega_1\ldots\omega_d.
    \end{aligned}
    \end{equation}
    Subsequently, the polynomials can be identified by Viete's formula as coefficients (up to a sign) of the characteristic polynomial of $H(\vec\lambda)$:
    \begin{equation}
        \det(H(\vec\lambda)-t\mathbb{1}) = \prod_{n=1}^d(\omega_n-t)=\sum_{n=0}^d c_{d-n}(\vec\omega) (-t)^n,
    \end{equation}
    but in this form it is obvious that each of the coefficients is simultaneously a polynomial in $\vec\lambda$. Therefore, ${\bf F}(\vec\omega(\vec\lambda))$ is a polynomial in $\vec\lambda$ as well.\\
    Since $F_{\vec m}$ is by construction a factor of $\bf F$, and by assumption vanishes on open subset $S$ of $\mathbb{R}^K$, the polynomial ${\bf F}(\vec \omega(\lambda))$ must also vanish everywhere.\footnote{Identity theorem, a well known result: Assume $D$ is an open and connected subset of $\mathbb{R}^K$, possibly the entire space. If two analytic functions $f, g$ agree on a an nonempty open set $S\subset D$, the functions are equal in entire $D$.}

    In particular, if we fix another $S'\in \Sigma$, ${\bf F}(\vec\omega(\vec\lambda'))=0$ for all $\vec\lambda' \in S'$. Since $\bf F$ is a product of a finite number of factors, each real analytic on $S'$ (on regions free of eigenvalue crossings the eigenvalues are simple roots of the characteristic polynomial and depend real-analytically on the parameters~\cite{kato_perturbation_1995}), at least one factor must be identically zero on $S'$. For a proof, assume the contrary: if no factor vanished identically on $S'$, each factor's zero set would be closed with empty interior\footnote{If one factor's zero set wasn't closed with empty interior, it would contain a ball, and by the identity theorem it would extent to the entire $S'$ -- contradiction with the assumption that no factor vanishes identically.}, and a finite union of such sets cannot cover the open set $S'$, contradicting that the product vanishes everywhere on $S'$.

    The factor that is identically zero on $S'$ has a form $\sum_{n=1}^d \omega_n(\vec\lambda') m_{\sigma(n)}$, therefore there exists $\vec m'$ that leads to rational dependence on $S'$ which was to be proven. It is also in $M_{\mathbb{Z}}$, since it's just permuted $\vec m$.
\end{proof}

Naturally, if an integer $\vec m$ showing rational energy dependence exists, it can be reinterpreted as a real vector, so if $\vec m$ shows dependence on $S$, there also exists a real vector $\vec x\in M_{\mathbb{R}}$ such that $\sum_n x_n \omega_n=0$. One of such vectors is $\vec x :=\vec m$. But this does not work in the opposite direction: a real energy dependence on a subset does not necessarily mean rational energy dependence.

What does work is contraposition: if we sample many points $\vec\lambda^{(1)},\ldots,\vec\lambda^{(N)}$ from some $S\in \Sigma$, and see that the matrix 
\begin{equation}L:=\begin{pmatrix}\vec \omega(\vec\lambda^{(1)})^T\\\vdots\\\vec\omega(\vec\lambda^{(N)})^T\end{pmatrix},\end{equation}
has zero-dimensional null space, or one spanned only by the excluded $(1,1,\ldots,1)$ vector, no real energy dependence exists, so no rational energy dependence could exist on any open subset $S$ of $\mathbb{R}^K$.

If we have numerically ascertained this, and the spectral criterion optimizer has found a satisfactory optimum at $\vec\lambda^*$ without degeneracies, it is still possible that there exists a rational energy dependence of $H(\vec\lambda^*)$. But we have just proven that for any $\vec m \in M_{\mathbb{Z}}$, the set of $\vec\lambda$ (constrained to the crossing-avoiding parameter subset $\bigcup \Sigma$) that give rise to rational eigenvalue dependence
\begin{equation}
    R_{\vec m} = \{\vec \lambda \in \bigcup \Sigma  : \sum \omega_n (\vec\lambda)m_n =0\},
\end{equation}
does not contain an open set -- if it did, it would extend to any $S'\in\Sigma$, and it would be detected as real eigenvalue linear dependence. Furthermore, it is closed in any $S\in\Sigma$, as $F(\vec \omega(\vec\lambda))$ is continuous on the set  $\cup_{S\in\Sigma} S$ (and preimages of closed sets under continuous maps are closed\footnote{The continuous map meant here is $\vec\lambda\mapsto \sum \omega_n (\vec\lambda)m_n$. The singleton $\{0\}$ is closed as a subset of $\mathbb{R}$, and its preimage must be therefore closed.}). There are countably many $\vec m$ in $M_\mathbb{Z}$, and a countable union of closed sets with empty interior cannot contain an open set, by the Baire category theorem. Therefore, in arbitrary small $\varepsilon$-vicinity of any $\vec\lambda^*$ we will find $\vec\lambda'$ that has no rational dependence. 

Since the torus sets $\mathcal T_{H(\vec\lambda'),\ket{i}}$ defined in Eq. \eqref{eq:torusstates} depend continuously on $\vec\lambda'$ away from crossings via the populations $p_n$ and eigenvectors, every state at  $\mathcal{T}_{H(\vec\lambda^*),\ket{i}}$ lies within a a distance $\delta$ to $\mathcal T_{H(\vec\lambda'),\ket{i}}$, and $\delta\rightarrow 0$ when $\varepsilon\rightarrow 0$. Hence the target fidelity attained at $\vec\lambda^*$ is attained arbitrarily close by the rationally independent $H(\vec\lambda')$.

\section{Dynamical Lie algebras of random control sets}
\label{app:dla}
In this section we present a derivation for the minimal number of generators we need such that their dynamical Lie algebra (DLA) is complete. 
We first treat the Pauli pool which can be seen as a more physics inspired derivation of the results presented in \cite{Smith2025,Aguilar2024} and then the qudit pool which has not been considered to the best of our knowledge in the form we present here.
The underlying physical reason in both cases is the same and relies on the transformation behaviour of the generator pools under the anti-unitary time-reversal operation. 

\subsection{DLA completeness for Pauli strings} \label{app:DLA_Pauli}
\subsubsection{Lower Bounding the Number of Generators}

We are given the seed set of generators from the Pauli pool $\{ P_j = P(\vec{v}_j)\}_{j=1}^\ell$. 
Each Pauli string is parametrized by a binary vector $\vec{v}_j \in\mathbb{F}_2^{2n}$ via $P(\vec{v}) = \bigotimes_{k=1}^{n}i^{\vec{v}_x\cdot \vec{v}_z} X^{v_{x,k}}Z^{v_{z,k}}$, with the convention $P(\vec{v}) = P^\dagger(\vec{v})$.

The central question is whether the seed of generators with a minimum of $\ell_{\rm min}$ is sufficient such that all nested commutators, together with their linear combinations, span the full Lie algebra $\mathfrak{su}(2^n)$, i.e.\ such that any Hamiltonian evolution can be implemented. 
A first naive count suggests that $\ell$ must scale exponentially in $n$, since $\mathrm{dim}(\mathfrak{su}(d=2^n))=d^2-1$ grows exponentially with the Hilbert-space dimension.

One structural property of the Pauli algebra reduces this bound dramatically. Two Pauli strings either commute or anticommute. 
When they anticommute, $\{ P(\vec{v}),P(\vec{w})\}=0$, the commutator is proportional to their product:
\begin{equation}
[P(\vec{v}),P(\vec{w})] = 2P(\vec{v})P(\vec{w}).
\end{equation}
The product of two (anticommuting) Pauli strings is itself a Pauli string, with binary-vector label given by the sum: $P(\vec{v})P(\vec{w})= i^{\vec{v}^\top G\vec{w}}P(\vec{v} +\vec{w})$ where $G=\sigma^+ \otimes \mathbb{1}_n$ and $\sigma^+=\sigma_x+i\sigma_y$. 
Commutation therefore generates precisely the Pauli strings labelled by $\mathbb{F}_2$-linear combinations of $\{\vec{v}_j\}_{j=1}^{\ell}$. 
If the $\vec{v}_j$ are linearly independent and $\ell = 2n$, they form a basis of $\mathbb{F}_2^{2n}$ and, provided all pairs anticommute, every Pauli string is reachable. 
The seed set then scales only as $\ell_{\rm min} = 2n = 2\log_2(d) = \mathrm{polylog}(d)$, exponentially smaller than the naive $O(\mathrm{poly}(d))$ bound.

\subsubsection{$\mathcal{T}$-odd obstruction}
However closer inspection reveals that $2n$ generators do not suffice. 
Every Pauli string obtained from the seed set by nested commutators transforms identically under the time-reversal operator $\mathcal{T} = U\mathcal{K}$, where $U = P(\vec{u})$ is a Pauli string and $\mathcal{K}$ denotes complex conjugation with $\mathcal{K^2}=1$.
For any choice of seed set there exists a $\vec{u}$, the unique solution of $2n$ linear equations, such that every seed element is $\mathcal{T}$-odd:
\begin{equation}
\mathcal{T}P_j\mathcal{T}^{-1} = - P_j \qquad j=1,\ldots,2n.
\end{equation}
This property is preserved under commutation: if $\mathcal{T}P\mathcal{T}^{-1}=-P$ and $\mathcal{T}Q\mathcal{T}^{-1}=-Q$, then the commutator $W=i[P,Q]$ is likewise $\mathcal{T}$-odd, $\mathcal{T}W\mathcal{T}^{-1} = -W$. The DLA grown from $\mathcal{T}$-odd seeds is therefore entirely $\mathcal{T}$-odd and cannot span $\mathfrak{su}(2^n)$ as we will show in the following.

The algebra in which the obstruction pins us is quantified by the square of the time-reversal operator:
\begin{equation}
\mathcal{T}^2 = UU^*= (-1)^{Y(U)} = \pm 1
\end{equation}
where $Y(U)$ counts the number of Pauli-$Y$ factors, since $\mathcal{K}Y\mathcal{K}=-Y$. 
The sign of $\mathcal{T}^2$ depends on the seed set and distinguishes the two inequivalent flavours of $\mathcal{T}$-odd can have; this then has consequences on the obstruction of the algebra.

\subsubsection{The $\mathcal{T}^2 = +1$ case and $\mathfrak{so}(2^n)$}

Consider the case $\mathcal{T}^2 = 1$, i.e.\ an even number of Pauli-$Y$ factors in $U$. There exists a unitary $V$ that reduces $\mathcal{T}$ to pure complex conjugation:
\begin{equation}
\mathcal{T} \rightarrow \tilde{\mathcal{T}} = V\mathcal{T}V^\dagger = \mathcal{K}.
\end{equation}
Before we move on let us briefly linger on why $\mathcal{T}=\mathcal{K}$ is referred to as time reversal operation \cite{wigner1932zeitumkehr}. 
Let us consider this operator acting on the position $x$ and momentum $p$ operators in real space. 
Under time reversal we expect the position operator to be invariant and indeed $\mathcal{T}x\mathcal{T}=x$. 
The momentum operator flips sign $\mathcal{T}p\mathcal{T}=-p$ since time reversal the sign of the velocity of a particle changes.

The above required transformation to simplify the time reversal operator is $V=\frac{e^{-i\pi/4}}{\sqrt{2}}(\mathbb{1}+iU)$ is a Clifford unitary: it sends Pauli strings to Pauli strings. 
Consequently, every element of the DLA remains a Pauli string, and the generator Hamiltonian $H_{\mathrm{DLA}}$ (a sum of DLA elements) satisfies $\mathcal{K}H_{\mathrm{DLA}}\mathcal{K} = -H_{\mathrm{DLA}}$.

Combining this $\mathcal{K}$-oddness with Hermiticity, $H_{\mathrm{DLA}} = H_{\mathrm{DLA}}^\dagger$, forces $H_{\mathrm{DLA}}$ to be antisymmetric as a matrix:
\begin{equation*}
H_{\mathrm{DLA}} = (H_{\mathrm{DLA}}^*)^\top = (-H_{\mathrm{DLA}})^\top = -H_{\mathrm{DLA}}^\top.
\end{equation*}
The unitaries $U = e^{-iH_{\mathrm{DLA}}}$ generated by antisymmetric matrices form the special orthogonal group $SO(2^n)$, and the Hamiltonians span the Lie algebra $\mathfrak{so}(2^n) \subsetneq \mathfrak{su}(2^n)$. The set of $\ell = 2n$ Pauli strings, together with all their commutators, therefore cannot generate arbitrary Hamiltonian evolutions.

The DLA of $\mathcal{T}$-odd Pauli strings contains precisely those Paulis with an odd number of $Y$ factors, $\mathcal{K}P\mathcal{K} =(-1)^{Y(P)} P =- P$. To reach the full $\mathfrak{su}(2^n)$, one must add Pauli strings with an even $Y$-count. A single such string suffices. 
Let $S$ be a Pauli string with an even number of $Y$ operators, and let $R$ be an arbitrary target Pauli string, also with even $Y$-count. We show that $R$ is reachable from $S$ and the DLA of odd-$Y$ Paulis in at most two commutator steps.

\paragraph{Anticommuting case, $\{S,R\}=0$.}
The product $P \equiv RS$ has an odd $Y$-count and therefore lies in the DLA. Since $P$ anticommutes with $S$, one commutator suffices:
\begin{equation}
S \;\xrightarrow{[P,\;\cdot]}\; [P,S] = 2R.
\end{equation}

\paragraph{Commuting case, $[S,R]=0$.}
Choose a Pauli string $T$ in the DLA that anticommutes with both $R$ and $S$, and set $S' \equiv T$ or $S' \equiv TS$ (depending on which combination is even).
Two successive applications of commutators then yield:
\begin{eqnarray}
S \;\xrightarrow{[S'S,\;\cdot]}\; [S'S,\,S] = 2S',  \nonumber \\
S' \;\xrightarrow{[RS',\;\cdot]}\; [RS',\,S'] = 2R. \nonumber
\end{eqnarray}
In both cases the intermediate Pauli strings carry an odd $Y$-count and thus belong to $\mathfrak{so}(2^n)$. 
Adding a single generator outside completes the algebra to $\mathfrak{su}(2^n)$. 
Therefore DLA is full for the minimal number of generators
\begin{equation}
\ell_{\rm min} = 2n + 1 = \mathrm{polylog}(d),
\end{equation}
which is an exponential improvement over the naive $\mathrm{poly}(d)$ bound.

Mathematically the phenomenon that a subalgebra is "suddenly" completed by adding another element outside it is studied in terms of the Cartan decompositions. 
The Lie algebra of the special unitary group can be decomposed in terms of 
\begin{equation}
\mathfrak{su}(2^n) = \mathfrak{so}(2^n) \oplus \mathfrak{r},
\end{equation}
where $\mathfrak{r} = \{P: Y(P)=\mathrm{even}\}$. 
This is the tangent space of the symmetric space $SU(2^n)/SO(2^n)$. 
This comes up math in the study of symmetric spaces \cite{zirnbauer1996riemannian,nakahara2003geometry} and in physics in terms of the Altland-Zirnbauer classification of random matrices \cite{altland1997nonstandard} and topological insulators where this decomposition is labeled $\mathrm{AI}$ \cite{schnyder2008classification}.

\subsubsection{$\mathcal{T}^2 =-1$ and completing $\mathfrak{usp}(2^n)$}

The discussion of $\mathcal{T}^2=-1$ is technically more challenging but proceeds accordingly. 
In this case we can find a basis $V$ in which the time reversal becomes
\begin{equation}
V\mathcal{T}V^\dagger = Y\otimes \mathbb{1}_{n-1} \mathcal{K}.
\end{equation}
This is the time reversal operation for spinful particles which in the case of real space flips both spin and reverses the sign of the momentum operator $\mathcal{T}\vec{\sigma} \mathcal{T}^{-1} = -\vec{\sigma}$ and $\mathcal{T}p \mathcal{T}^{-1} = -p$.
There encounter the Cartan decomposition $\mathfrak{su}(2^n)=\mathfrak{usp}(2^n) \oplus \mathfrak{r}$ of type AII. 
Here the remainder $\mathfrak{r}$ are Pauli matrices where the first Pauli is $\mathbb{1}$ and the rest of the Pauli string contains an even number of $Y$-Paulis. 
For the first Pauli different than the identity, the rest of the Pauli string has to contain an odd number of $Y$-Paulis.

\subsection{DLA Completeness for a Qudit} \label{app:DLA_qudit}

Consider a $d$-level qudit with basis $\{\vert 1\rangle,\ldots,\vert d\rangle\}$. 
We are given $\ell$ generators drawn from two families: population-reshuffling operators $X_{ij}$, $Y_{ij}$ that exchange amplitude between $\vert i\rangle$ and $\vert j\rangle$, and phase operators $Z_i$ that shift the relative phase between $\vert i\rangle$ and $\vert i{+}1\rangle$ (see Eq.~\eqref{eq:quditpool}). 
The question is the minimal $\ell_{\rm min}$ for which the DLA of the seed set spans the full Lie algebra $\mathfrak{su}(d)$ i.e. any Hamiltonian can be reached.

\subsubsection{Population reshuffling algebra}
To treat $X$- and $Y$-type generators on the same footing, we write the general population-reshuffling operator
\begin{equation}
h_{jk}(\alpha) = e^{i\alpha}\vert j\rangle \langle k \vert + \mathrm{h.c.}
\end{equation}
where $\alpha = 0$ gives the $X$-type and $\alpha = -\pi/2$ the $Y$-type. The commutator of two reshufflings on adjacent bonds $i \leftrightarrow j$ and $j \leftrightarrow k$ yields a previously uncoupled pair:
\begin{equation}
[h_{ij}(\alpha),h_{jk}(\beta)] = ih_{ik}(\alpha +\beta - \pi/2).
\end{equation}
This relation implies that populations in $\vert i\rangle$ and $\vert f\rangle$ 
can be reshuffled by successive commutators provided there exists a chain of generators connecting them
\begin{equation}
\vert i \rangle \overset{h_{ia_1}}{\longrightarrow} \vert a_1\rangle \longrightarrow \ldots \longrightarrow \vert a_n \rangle \overset{h_{a_nf}}{\longrightarrow} \vert f\rangle,
\end{equation}
with phases $\alpha \in \{0,\,\pi/2\}$.

\subsubsection{Graph interpretation}
The connectivity structure is captured by a graph introduced in \cite{Altafini2002,SchirmerFuSolomon2001}, with vertex set $V=\{1,\ldots,d\}$ with the labels of the states.  
The edge set $E=\{ (i,j): i\ne j \land i,j \in V\}$ coincides with the set of population reshuffles in the minimal seed set. 
The minimal connected subgraph is a spanning tree: It contains a set of $d-1$ edges with no loops, providing a unique path $j \overset{\gamma}{\rightarrow}k$ between any two vertices. 
Successive commutators along the path $\gamma$ yield
\begin{equation}
h_{jk}\!\left(\textstyle\sum_{e\in\gamma} \alpha_e - (n-1)\tfrac{\pi}{2}\right).
\end{equation}
The spanning tree therefore requires $\ell = d-1$ population-reshuffling generators.

\subsubsection{The $\mathcal{T}$-odd obstruction}

The $d-1$ generators of a spanning tree do not span the entire $\mathfrak{su}(d)$. As in the qubit case, the obstruction is a time-reversal symmetry $\mathcal{T}=U\mathcal{K}$ with $U=\sum_je^{i\varphi_j}\vert j\rangle \langle j \vert$. 
This obstruction was discussed more formally in \cite{Schirmer2002} without connections to the time reversal operator.
Applying $\mathcal{T}$ from both sides, we can move the unitary past the complex conjugations which gives $\mathcal{T}h_{jk}\mathcal{T}^{-1} = \mathcal{K}U^\dagger h_{jk} U \mathcal{K} $ for any generator. 
For any spanning tree there exists a unitary $U$ such
\begin{equation}
U^\dagger h_{jk}(\alpha_{jk})U = h_{jk}(-\pi/2) \qquad (j,k) \in E,
\end{equation}
so that in the basis defined by $U$ every generator is $\mathcal{K}$-odd:
\begin{equation}
\mathcal{K}h_{jk}(-\pi/2)\mathcal{K} = -h_{jk}(-\pi/2).
\end{equation}
The phases $\{\varphi_j\}$ are fixed by the following procedure: pick an arbitrary root vertex $i$ and set $\varphi_i = 0$; for each neighbouring vertex $j$ connected by an edge, set $\varphi_j = \varphi_i + \alpha_{ji} - \pi/2$; repeat until all vertices are assigned. The spanning-tree structure guarantees that this is always consistent (no loop constraints).

Since the commutator of two $\mathcal{K}$-odd operators is again $\mathcal{K}$-odd, the entire DLA is $\mathcal{K}$-odd. The generators take the form $H_{\mathrm{DLA}}=\sum_{i<j}\omega_{ij}Y_{ij}$, the Hamiltonians are antisymmetric, and the unitaries they generate form $SO(d)$ \cite{schirmerleahy2002,albertini_notions_2003}. The DLA is $\mathfrak{so}(d) \subsetneq \mathfrak{su}(d)$.

In physics this phenomenon is known as a\textit{pure gauge} and it is the reason why on trees and especially on one-dimensional lattices, magnetic fields, which break time reversal invariance, can be gauged away. 
This has the physical consequence that on trees or simply connected spaces there is no Aharonov-Bohm effect and also no quantum Hall effect in one dimension \cite{schnyder2008classification}. 

\subsubsection{Completing the DLA}

We will now show, that again, a single additional population-reshuffling generator $h_{kl}(\alpha_{kl})$ can suffice to complete the DLA.
Because the original $d-1$ generators form a spanning tree, adding one more edge creates exactly one loop. For any two vertices $i,j$ on that loop there are two distinct paths, $\gamma_1$ and $\gamma_2$, yielding two reshufflings $h_{ij}(\alpha_{\gamma_1})$ and $h_{ij}(\alpha_{\gamma_2})$ with different phases. Their commutator produces a diagonal (phase) operator:
\begin{equation*}
[h_{jk}(\alpha),h_{jk}(\beta)] = 2i \sin(\alpha-\beta)\,(\vert j\rangle\langle j\vert - \vert k \rangle \langle k\vert)
\end{equation*}
If this or more generally, any diagonal operator $D=\sum_{j}d_j\vert j \rangle \langle j\vert$, acts on a reshuffling generator it rotates its phase by $\pi/2$:
\begin{equation}
[D,h_{jk}(\alpha)]=i(d_j-d_k)h_{jk}(\alpha-\pi/2).
\end{equation}
Thus $D$ converts $Y$-type generators into $X$-type generators (and vice versa). 
Commuting the two types on the same bond then recovers the $Z$-type operators:
\begin{equation*}
[X_{j,j+1},Y_{j,j+1}]=[h_{j,j+1}(0),h_{j,j+1}(-\pi/2)]\sim Z_j.
\end{equation*}
The DLA of a spanning tree plus one such additional population-reshuffling generator spans the full Lie algebra $\mathfrak{su}(d)$. 
Here again we encountered the AI decomposition of the full algebra in terms of the orthogonal Lie algebra and the population reshufflings not on the spanning tree $\mathfrak{su}(d) = \mathfrak{so}(d) \oplus \{ h_{ij}(\alpha_{ij})\}_{(i,j)\notin E}$.
Altogether we find that the minimal seed size is
\begin{equation}
\ell_{\min} = d = \mathrm{poly}(d),
\end{equation}
linear in the Hilbert-space dimension. 
Remarkably, exponentially more generators are needed than in the case of Pauli strings.
This is attributed to the lack of algebraic structure of the qudit generators compared with the case of Pauli strings.

\section{Optimization procedure details}
\label{app:optimdetails}

Instead of maximizing $\mathcal{F}$ of Eq. \eqref{eq:specfid} directly, we try to find $\vec\lambda$, for which the vector of differences 
\begin{equation}
    \vec\delta=(\sqrt{p_{i,n}}-\sqrt{p_{f,n}})_n
\end{equation} is closest to zero using \verb|scipy.optimize.least_squares|. This function attempts to minimize the norm of this vector using all of the coordinates of $\vec\delta$, which is better behaved numerically than direct optimization of the scalar $\mathcal{F}$, but mathematically is  in fact equivalent to it, as
\begin{equation}\begin{aligned}
    \sum(\sqrt{p_{i,n}}-\sqrt{p_{f,n}})^2 &= 2-2\sum_{n}\sqrt{p_{i,n}p_{f,n}}\\&=2(1-\mathcal{F}).\end{aligned}
\end{equation}

The optimizer is local, so we restart it from 1000 random parameters $\vec\lambda$; target is declared unreachable if the above optimization procedure yields maximum fidelity below $1-10^{-10}$. If above the threshold, is is declared (approximately) reachable. This budget suffices to declare explicitly reachable states $e^{-i\sum_\ell\lambda_\ell h_\ell}\ket{i}$ as reachable with probability larger than $95\%$ for the cases mentioned in Section \ref{sec:breakdown}. \footnote{As an additional check -- to not rely on 3 qubit data only -- we repeated the analysis with GUE generators at low $d$: for $K<d$ no Haar targets are found reachable, while explicitly constructed reachable states are.} For the final results of Sec. \ref{sec:numericalspectral} involving $d=32$ 5-qubit and qudit pools, we employ a variable number of reruns with maximum of $10^5$ initial $\vec\lambda$ probed -- the computational complexity is so high that we can only declare states as approximately reachable by much looser threshold of $\mathcal{F}\ge0.99$, and strict comparisons are impossible in this case.

Initial formulation as population  matching (subsequent fidelity fine tuning doesn't affect this) sheds light on the complexity of the problem: we have to match $p_{i,n}=p_{f,n}$, both vectors in $\mathbb{R}^d$, while having $K$ parameters $\vec\lambda$ to change. Due to normalization of the probability vectors and their invariance under rescaling of $\vec\lambda$, in fact we end up with $d-1$ equations in $K-1$ variables. 

If these were \emph{generic} equations, they would have no solutions for $K<d$ parameters -- they would be overconstrained. This leads to optimization problems even when the target state is explicitly constructed as reachable. This problem is manageable for 3 qubits ($d=8$), but becomes fully numerically unmanageable for 4 qubits ($d=16$). 

At $K=d$ generically isolated points are solutions, and they become easier to find. At $K>d$, generically there are entire continous families of parameters $\vec\lambda$, for which $\ket{f}$ is reachable from $\ket{i}$ with $H(\vec\lambda)$, provided it is reachable at all. This is also reflected in the numerical data: optimization becomes faster and more reliable.

Formally speaking, we are able to numerically suggest that in some cases (low dimensions and all $K$; or high dimensions with $K\ge d$) explicitly reachable states are often found reachable. However, this does not tell us much about the optimization quality of \emph{random} states which were \emph{not} found reachable. In either case, even if there are reasons to suspect that optimization was not exhaustive, from each run we still get the best found fidelity $\mathcal F_*$, which \emph{is} a lower bound to the theoretically maximum value. 

\begin{figure*}[!t]
    \centering
    \includegraphics[width=1\linewidth]{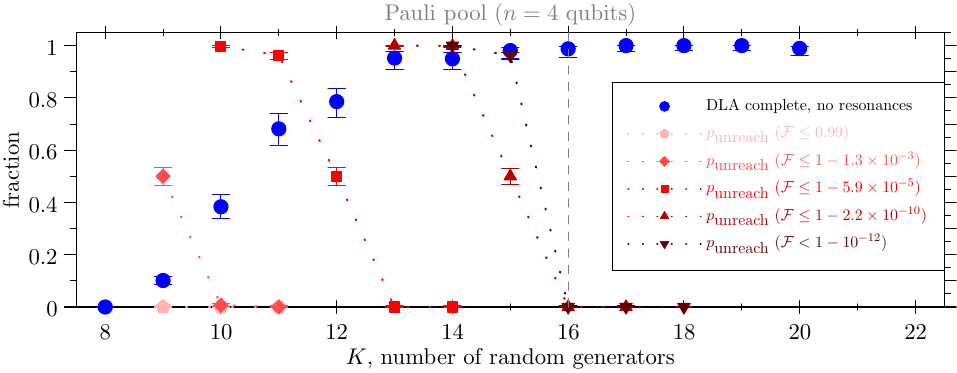}
    \caption{%
    Approximate reachability of the Pauli pool at $d=16$ (four qubits). For each number of generators $K$, control sets of $K$ operators and uniformly random states are sampled, and the spectral overlap between initial and final state is optimized. A state is deemed unreachable if the optimized overlap is below a prescribed threshold, $\mathcal{F}_*<1-\varepsilon$; otherwise it is \emph{approximately reachable}. We plot the probability of unreachability $p_{\text{unreach}}(K)$ at different thresholds; discrete points corresponding to the same threshold are connected with a dotted line to guide the eye. For $K<d$, we know this procedure to not be exhaustive -- it is almost impossible to find the true optimum (and e.g. explicitly reachable states are not found as reachable with strict thresholds). However, this procedure still gives us an upper bound to $p_{\text{unreach}}$, as it is impossible to misclassify a state as approximately reachable. \\
    Notably, a large fraction of states is \emph{approximately reachable} even for tiny threshold. At $K=9$ -- the minimal number of generators leading to full dynamical Lie algebra (see Sec. \ref{sec:dla}) -- at least half of the state space is approximately reachable with fidelity of $1-1.3\cdot 10^{-3}$ with random control sets.
    \label{fig:fourqubitdata}}
\end{figure*}

We can employ this observation in order to show that for Pauli pool on $4$ qubits, \emph{most of the state space sits very close to the reachable set} even for very low $K$, below the $K=d$ threshold of Section \ref{sec:theory} -- see Fig. \ref{fig:fourqubitdata}. This is not contradictory, as the threshold is looser, and low-dimensional set gets inflated this way. 

\section{Proof of Proposition \ref{prop:cutoff}} \label{app:cutoffproof}
The proof is constructive: given enough control terms $(h_1,\ldots,h_K)$, and fixed initial $\ket{i}$ and final states $\ket{f}$, we will always find weights $(\lambda_1,\ldots,\lambda_K)$ {in $V_*$} such that the relevant quadratic form (Eq. \eqref{eq:diffQ}) is nonnegative, and another weights such that it is nonpositive. Therefore, it can not be strictly positive or negative definite. We exclude the case when $\ket{f}\propto\ket{i}$, as it is irrelevant and non-generic.

\begin{proof} 
Let $\vert i\rangle$ and $\vert f\rangle$ span a two-dimensional subspace $W$; let $P$ be the projector onto it. 
For a state $\vert\psi\rangle\in W$, splitting $H\vert\psi\rangle$ into its components inside and outside $W$ gives
\begin{equation}
\langle\psi\vert H^2\vert\psi\rangle=\langle\psi\vert h^2\vert\psi\rangle+\Vert(\mathbb{1}-P)H\vert\psi\rangle\Vert^2 ,
\label{eq:split}
\end{equation}
where $H=H(\vec{\lambda})$ and $h=PHP|_W$ is $H$ restricted to $W$, a $2\times2$ matrix. 
Suppose a $\vec\lambda$ exists such that $\langle H(\vec\lambda)\rangle_i=\langle H(\vec\lambda)\rangle_f$ (that is, $\vec\lambda\in V_*$) and additionally $(1-P)H\ket{f}$ vanishes, so the last part of Eq. \eqref{eq:split} vanishes for $\ket{\psi}=\ket{f}$. We first show that such a $\vec\lambda$ yields $\langle f|H^2|f\rangle\le\langle i|H^2|i\rangle$ (and symmetrically that $\langle f|H^2|f\rangle\ge\langle i|H^2|i\rangle$ for other $\vec\lambda$), and prove its existence at the end. 

Since $W$ is two-dimensional, $h$ has two eigenvectors $\vert a\rangle$, $\vert b\rangle$ with eigenvalues $a$, $b$. Writing $\vert\psi\rangle=c_a\vert a\rangle+c_b\vert b\rangle$ with $\vert c_a\vert^2+\vert c_b\vert^2=1$, we find
\begin{align}
\langle\psi\vert h\vert\psi\rangle&=a\vert c_a\vert^2+b\vert c_b\vert^2 ,\\
\langle\psi\vert h^2\vert\psi\rangle&=a^2\vert c_a\vert^2+b^2\vert c_b\vert^2 \nonumber\\
&=(a+b)\langle\psi\vert h\vert\psi\rangle-ab ,
\end{align}
so the second moment of $h$ is fixed by its mean. But $\langle\psi\vert h\vert\psi\rangle=\langle\psi\vert H\vert\psi\rangle$ for states in $W$, and in particular $\ket\psi=\ket i,\ket f$, and we have picked $\vec\lambda$ such that the means are equal. Then Eq.~\eqref{eq:split} gives
\begin{equation}
\langle f\vert H^2\vert f\rangle=\langle f\vert h^2\vert f\rangle=\langle i\vert h^2\vert i\rangle\le\langle i\vert H^2\vert i\rangle ,
\end{equation}
where the inequality holds because the second term in Eq.~\eqref{eq:split} is non-negative for $\vert i\rangle$.

By the definition of $\Delta Q$, $\vec{\lambda}^\top\Delta Q\,\vec{\lambda}=\langle f\vert H^2\vert f\rangle-\langle i\vert H^2\vert i\rangle\le 0$, so $\Delta Q\vert_{V_*}$ is not positive definite. Exactly the same reasoning, with $\ket i$ and $\ket f$ swapped, proves that (if a $\vec\lambda\in V_*$ such that $(1-P)H\ket{i}=0$ can be found) $\Delta Q$ is not negative definite either.

It remains to show that such a $\vec{\lambda}\neq\vec 0$ exists. Consider, without loss of generality, a basis in which
\begin{equation}
    \ket{f}=\begin{pmatrix}1\\0\\0\\\vdots\\0\end{pmatrix},~~ \ket{i}=\begin{pmatrix}u\\v\\0\\\vdots\\0\end{pmatrix},
\end{equation}
and construct a rectangular $d\times 3$ matrix with colums made of $\ket{f},~\ket{i}$ and $H(\vec\lambda)\ket{f}$:
\begin{equation}
    \left(\begin{array}{c|c|c}
    \begin{matrix}1\\0\\0\\\vdots\\0\end{matrix}
    &
    \begin{matrix}u\\v\\0\\\vdots\\0\end{matrix}
    &
    H(\vec\lambda)\ket{f}
    \end{array}\right).
\end{equation}
This matrix has rank 2 -- and thus, $H(\vec\lambda)\ket{f}$ lies in the span of $\ket{f}$ and $\ket{i}$ -- if and only if last $d-2$ coefficients of $H(\vec\lambda)\ket{f}$ in this basis are all zero. This gives $2(d-2)$ linear equations in the real coefficients of $\vec\lambda$, as both real and imaginary parts have to be fixed. Together with the linear requirement that $\langle H\rangle _i = \langle H\rangle_f$, this gives $2(d-2)+1=2d-3$ real homogenous equations in $K$ variables; this problem always has nonzero solutions if $K\ge(2d-3)+1=2d-2$, which is what we find numerically.
\end{proof}
\end{document}